\documentstyle[12pt,graphicx,epsfig]{article}
\newcommand{\be}{\begin{eqnarray}}

\newcommand{\ee}{\end{eqnarray}}

\makeindex

\title{
         What Have We Learned from RHIC?\footnote
{Invited talk presented at the International
Conference on the Physics and Astrophysics of the 
Quark Gluon Plasma, Nov. 26-30 2001}}

\author{Larry McLerran
        \\
       {\small\it Nuclear Theory Group, Brookhaven National Laboratory,
        Upton, NY 11793  } \\      
       }

\begin{document}
\maketitle
\parindent=20pt

\begin{abstract}
In this talk, I present what I believe
we have learned from the recent RHIC heavy 
ion experiments.  The goal of these experiments is to make and study
matter at very high energy densities, greater than an order of magnitude
larger than that of nuclear matter.  Have we made such matter?  What have we
learned about the properties of this matter?  What do we hope and expect 
to learn in the future?

\end{abstract}

\section{Introduction}

The goal of the heavy ion program at the RHIC at Brookhaven National Laboratory
is to make and study new forms of matter at energy densities in 
excess of ten times
that of nuclear matter.  I will describe the status of this program 
from a theorist's
perspective.  In this talk, I would like to address four simple questions:
\begin{itemize}

\item{\bf What are we trying to understand?\\}

We are trying to produce new forms of matter and understand their
properties.

\item{\bf What have we already learned?\\}

I will argue that we have produced matter at energy densities at least 10 
and perhaps 100 times higher 
than that of nuclear matter.  
This matter is strongly interacting with itself.

\item{\bf What do we expect to learn?\\}

There are some measurements which will be carried out in the near term
which will answer specific questions about the properties of matter
produced in RHIC collisions.

\item{\bf What do we hope to learn?\\}

In the longer term, there
are the measurements which are harder to understand
or more controversial or ambiguous in their interpretation,
and will require new analysis.

\end{itemize}

\section{What Are We Trying to Understand?}

There are two central issues of the RHIC experimental program.
\begin{itemize}

\item{\bf What is the behavior of matter at  
asymptotically large  energy density? \\ }

Matter at very high energy density in thermal equilibrium is believed to
form a Quark Gluon Plasma.  This gas of almost free quarks and gluons
is thought to be the proper description of matter when energy densities
are larger than about $1 ~ GeV/Fm^3$. 
This is about the energy density inside a 
proton or neutron and is about an order of magnitude larger than 
that of nuclear matter.
Matter of this energy density occurs naturally 
in the cores of neutron stars and
was present during the big bang.

\item{\bf What type of  matter is important for high energy hadrons? \\ }

This matter  is believed to
be a very dense system of quarks and gluons, 
and is called a Color Glass Condensate.
It is an incoherent superposition of 
quantum mechanical states of Bose condensates.  
It is parameterized by a surface energy
density, and is presumably the correct description of matter 
when the energy density per unit area is greater than about $1~ GeV/Fm^2$
The Color Glass Condensate may be responsible 
for universal behavior of all hadronic
interactions at high energy.

\end{itemize}

\subsection{The Quark Gluon Plasma}

\begin{figure}[htb]
    \begin{center}
        \includegraphics[width=0.75\textwidth]{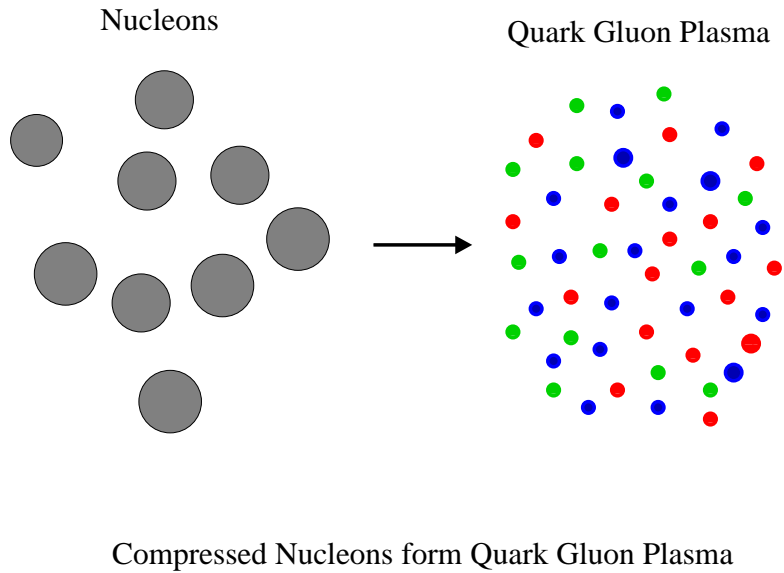}
        \caption{Nuclear matter when compressed 
becomes a gas of quarks and gluons: the Quark Gluon Plasma}\label{nqgp}
    \end{center}
\end{figure}

In Fig. \ref{nqgp},  a cartoon is shown which 
illustrates what we expect happens
when nucleons, or for that matter any hadrons, 
are compressed to densities higher
than is typical inside a nucleon.  Eventually the constituents of the
nucleon travel more or less freely around the system as a whole.  
This system is the Quark Gluon Plasma.

The formation of a Quark Gluon Plasma 
should begin to occur at about the energy density of matter
inside a proton.  Since a proton has a size about 
$R \sim 1~Fm$, and has a rest mass energy of
$M \sim 1~GeV$, corresponding to  an energy density of about $1~GeV/Fm^3$

\begin{figure}[htb]
    \begin{center}
        \includegraphics[width=0.75\textwidth]{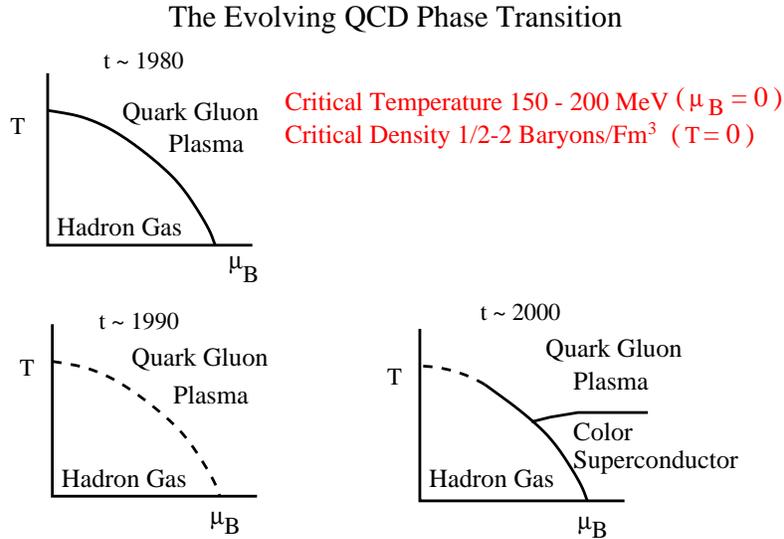}
        \caption{Beliefs about the QCD phase 
diagram as a function of time.  Solid line indicates phase boundary, and
dashed a rapid crossover.}\label{phd}
    \end{center}
\end{figure}

These observations led a number of people in the late 1970's to suggest that
there was a phase transition between ordinary matter and a Quark Gluon Plasma
as is shown in 
Fig. \ref{phd}.\cite{earlywork}
These conjectures were later firmed up by lattice Monte-Carlo 
simulations.\cite{earlymc}

The early lattice Monte-Carlo simulations were unable to properly include
the effects of quark masses.\cite{pd}  
About the mid 1980's arguments appeared which
strongly suggested that for realistic values of quark masses, a real phase 
transition does not occur.  Instead there is a sharp change in the properties
of the system at some temperature and baryon chemical potential.  Although it
does not have strong implications for heavy ion collisions, this would affect 
strongly cosmological scenarios and has potential effect in neutron star cores.

By 2000, the story had become more complicated.  It is now 
believed that there is a
line of first order phase transitions in the baryon number chemical potential 
and temperature plane.  There may be some region where the transition 
is second order or crossover.  
At  high baryon number density and small temperature, 
there may be a number of color superconducting phases.\cite{colorsuper}  

The bottom line on this historical aside is that physics is an experimental 
discipline.  I think it is unlikely we will have a first 
principles understanding
of the phase transition in QCD on the basis of pure thought.  Much work will
have to be done which combines experimental results and difficult numerical, 
presumably lattice, Monte-Carlo simulation, before we will have a compelling
picture of high density matter. 

At zero baryon number density, a lot has been learned about 
the properties of high
temperature matter.  For technical reasons, 
lattice Monte Carlo methods are very
difficult to implement at finite baryon number density.
At finite temperature, one has made remarkable progress.\cite{recentmc}
\begin{figure}[htb]
    \centering
       \mbox{{\epsfig{figure=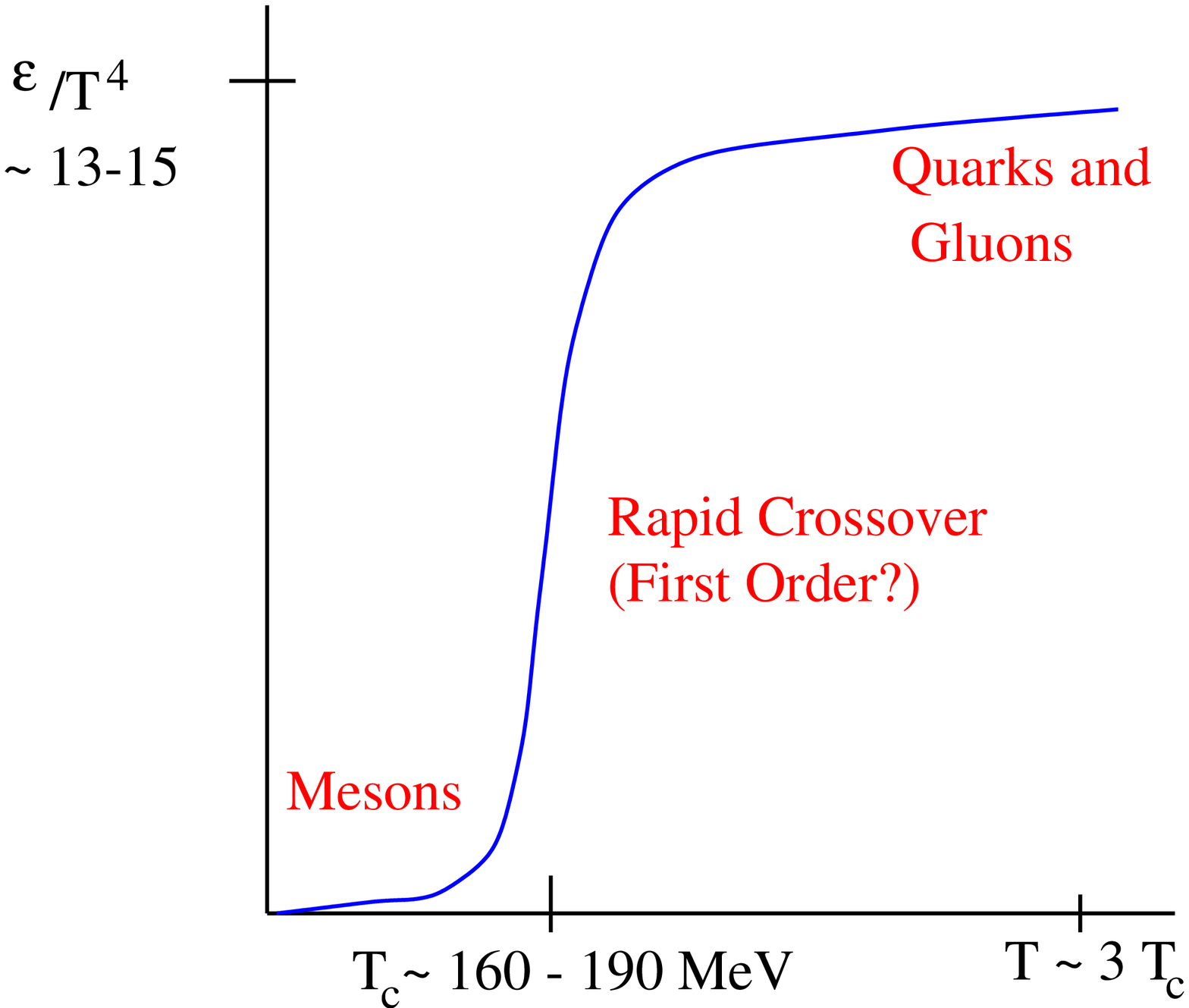,
        width=0.40\textwidth}}\quad
             {\epsfig{figure=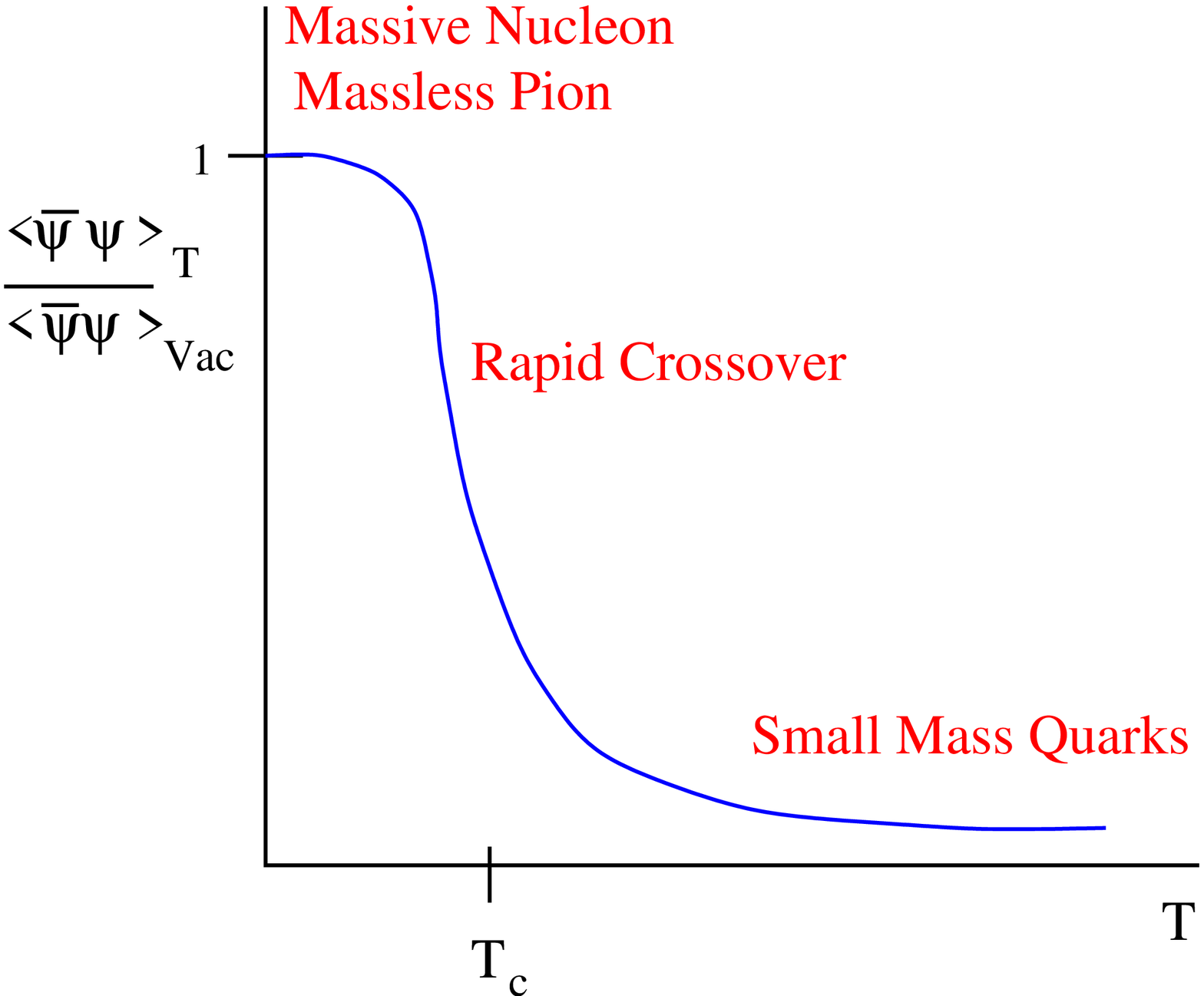,
        width=0.40\textwidth}}}
        \caption{(a)The energy density vs temperature. 
(b) The chiral condensate vs T.   
Both plots are at zero baryon number density. }
        \label{evst}
\end{figure}

In Fig. \ref{evst}a, a plot of the energy density, $\epsilon$ scaled by $T^4$
is shown vs temperature, $T$.  This variable $\epsilon/T^4$ is 
roughly proportional to the number of particle degrees of freedom important
at the energy scale corresponding to $T$.
A rapid cross over occurs at a temperature 
of $T \sim 160-190~MeV$.  A first order phase transition cannot be ruled out
at this time because the effects of finite quark mass are not under sufficient
control.  At low temperatures, the number of degrees of freedom are small
and consistent with a gas of pions.  At high temperatures, the degrees
of freedom are roughly that of a massless gas of quarks and gluons.  A sharp
transition occurs between the meson system and that of the quark and gluons 
where $\epsilon/T^4$ changes by an order of magnitude.  By a temperature
of order $T \sim 500 ~MeV$, the number of important particle degrees of 
freedom stays approximately constant.

The origin of the proton and neutron mass is one of the mysteries 
of QCD.  The masses of the up and down quarks inside a proton are only a 
few percent of the nucleon mass.  Do not be fooled by those who
say that the LHC Higgs boson search is 
designed to explain the origin of mass.  The LHC
probes electroweak physics and this gives masses to the up and down
quark masses, which can explain only 
a few percent of the nucleon mass.  Even in a world 
where the up and down quark masses had no mass and electroweak physics
was entirely ignored, the nucleon would still have roughly its present 
mass, and this arises from QCD.
It is believed that this mass is obtained by breaking of a chiral symmetry
of the strong interactions.  
\begin{figure}[htb]
    \centering
       \mbox{{\epsfig{figure=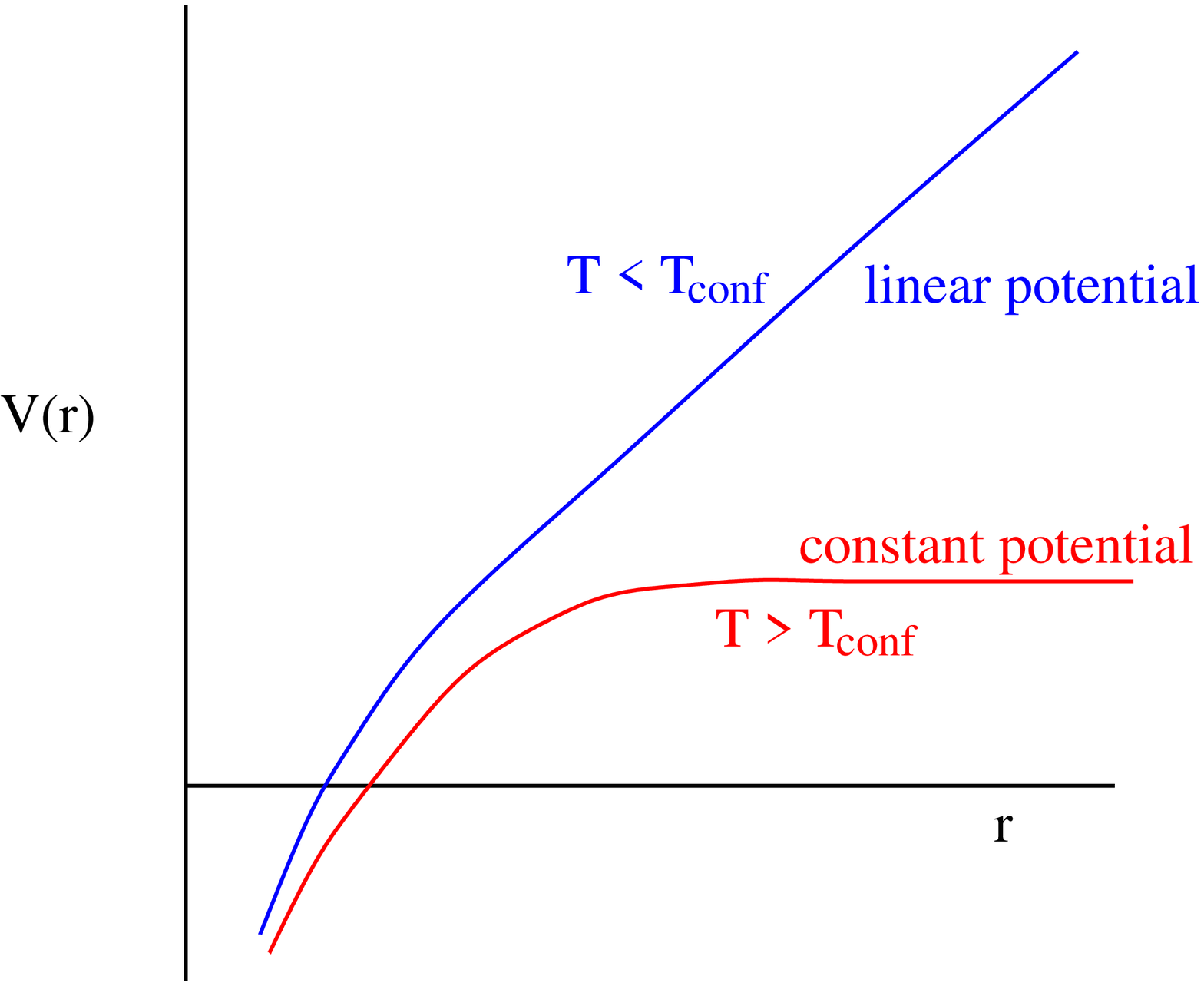,
        width=0.40\textwidth}}\quad
             {\epsfig{figure=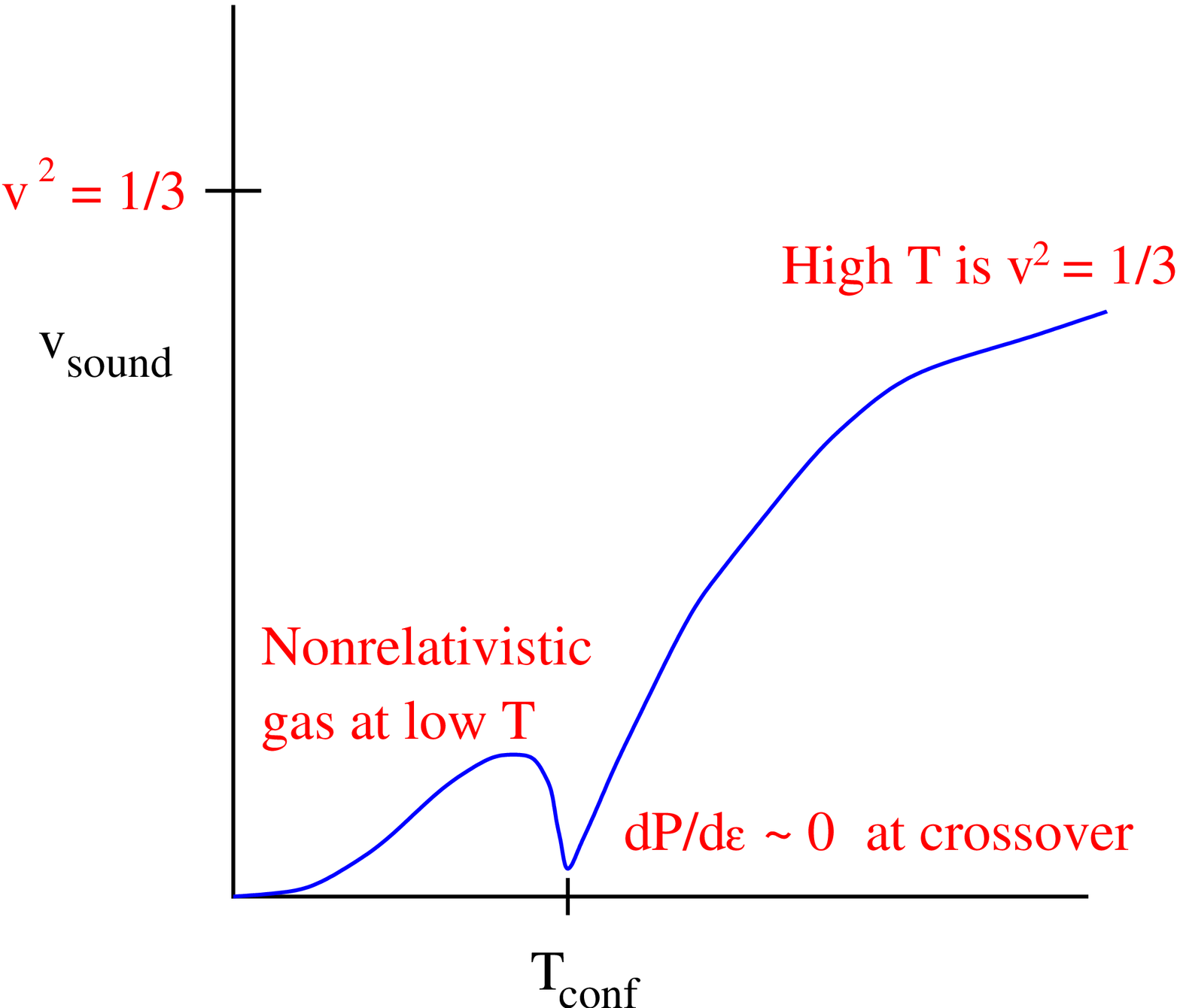,
        width=0.40\textwidth}}}
        \caption{(a)The potential between two quarks as a function
of distance. (b)The sound velocity as a function of temperature.   }
        \label{vr}
\end{figure}

This symmetry is broken in our low temperature world but is restored
in the high temperature world.  A measure of this symmetry breaking
is $<\overline \Psi \Psi >$, the typical value of a condensate.  This
condensate is composed of quark-antiquark pairs.
At low temperatures, this is non-zero, and rapidly goes to zero at the
phase transition temperature, as is seen in Fig. \ref{evst}b.
The nucleon mass is proportional to this condensate, and so it goes to zero
in the Quark Gluon Plasma.

Showing that the confining force between two quarks is linear 
was one of the first triumphs of
lattice QCD.  It can be measured as a function of temperature, and indeed
in the Quark Gluon Plasma it goes to a constant at long distances.  This is 
shown in Fig. \ref{vr}a.

How is this transition from a gas of pions into a Quark Gluon Plasma
manifest in terms of bulk properties of the system?  The
sound velocity is $v_s^2 = dP/d\epsilon$ where $P$ is the pressure.
At the crossover, the pressure is roughly constant while the energy density
changes by an order of magnitude.
The sound velocity should drop to near zero at $T_c$, as shown in 
Fig. \ref{vr}b.

\subsection{The Color Glass Condensate}

\begin{figure}[htb]
    \begin{center}
        \includegraphics[width=0.35\textwidth]{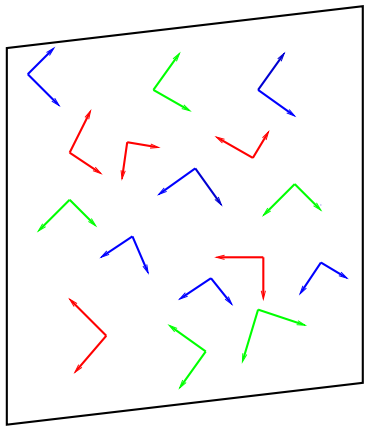}
        \caption{The Color Glass Condensate.  }\label{colorglass}
    \end{center}
\end{figure}

To understand this new form of matter, it is convenient to imagine
a hadron in a reference frame where it has very large longitudinal momentum.
We will be interested in the constituents of the hadron wavefunction 
which have  small longitudinal momentum in this frame of reference.
These low momentum constituents are produced by the high momentum ones.
Because the high momentum constituents appear to have time scales which
are Lorentz time dilated compared to their natural scales, and since they 
induce the low momentum fields associated with the low momentum 
particles, the low momentum fields evolve very slowly compared to their natural
time scale.  Hence the term, Color Glass, since the fields are composed of
color gluons, and glass because the time scale for evolution of these low 
momentum fields is much longer than their natural time scale.  
These fields live on a two dimensional sheet
because of the Lorentz contraction of the high energy hadron.  
We shall argue in the following
paragraphs that the phase space-density of these fields becomes large and forms
a condensate.\cite{cgc}

The fields on the two dimensional sheet turn out to be similar to the
Lienard-Wiechart potentials of electrodynamics.  They correspond to plane 
waves as in the Weizsacker-Williams approximation of 
electrodynamics, except that they have color.  
They have their color electric field perpendicular
to their color magnetic field and both perpendicular to their direction of 
motion, $ \vec{E}^a \perp \vec{B}^a \perp \vec{z}$.  
They have a random color.  This is shown in Fig. \ref{colorglass}

\begin{figure}[htb]
    \centering
       \mbox{{\epsfig{figure=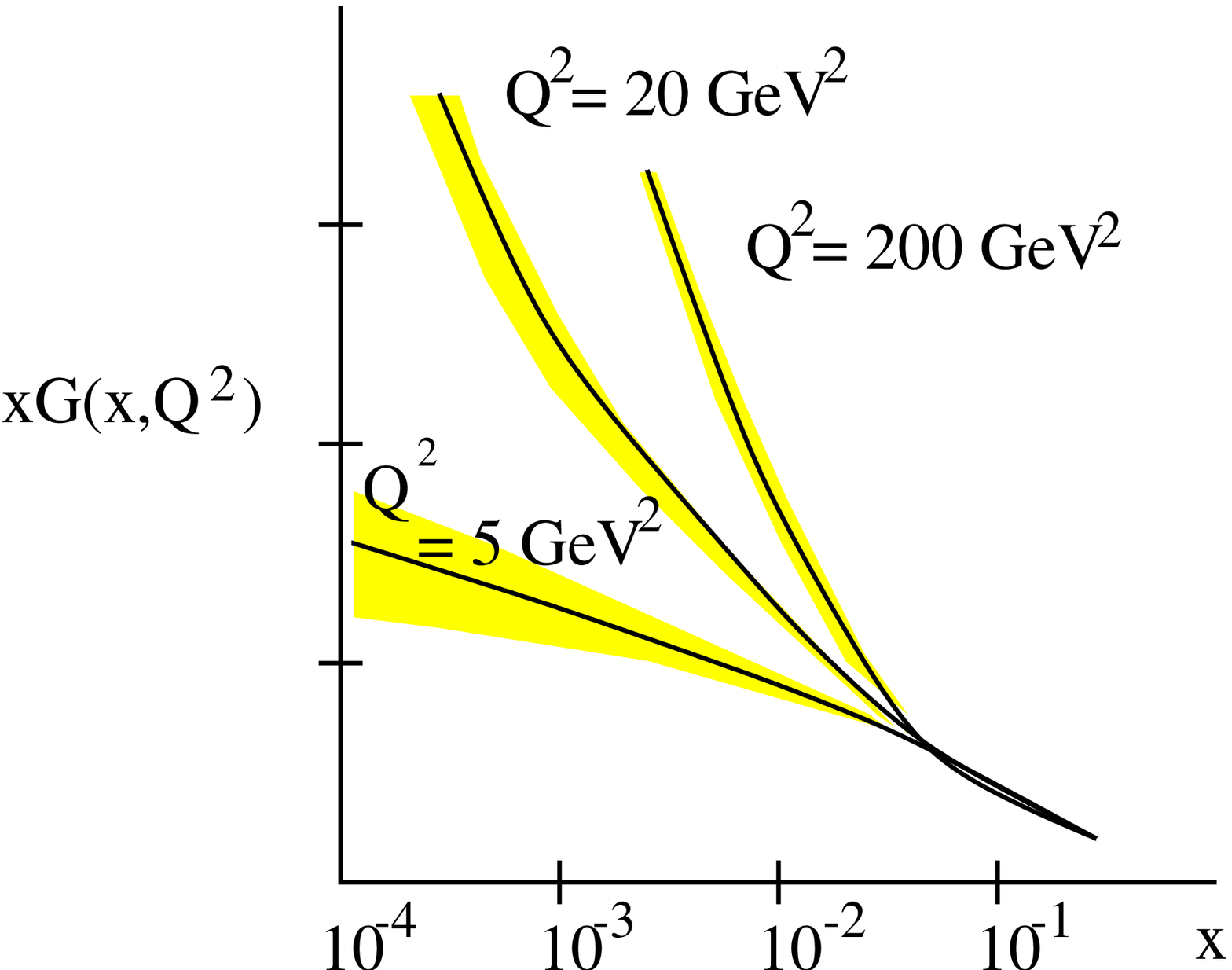,
        width=0.40\textwidth}}\quad
             {\epsfig{figure=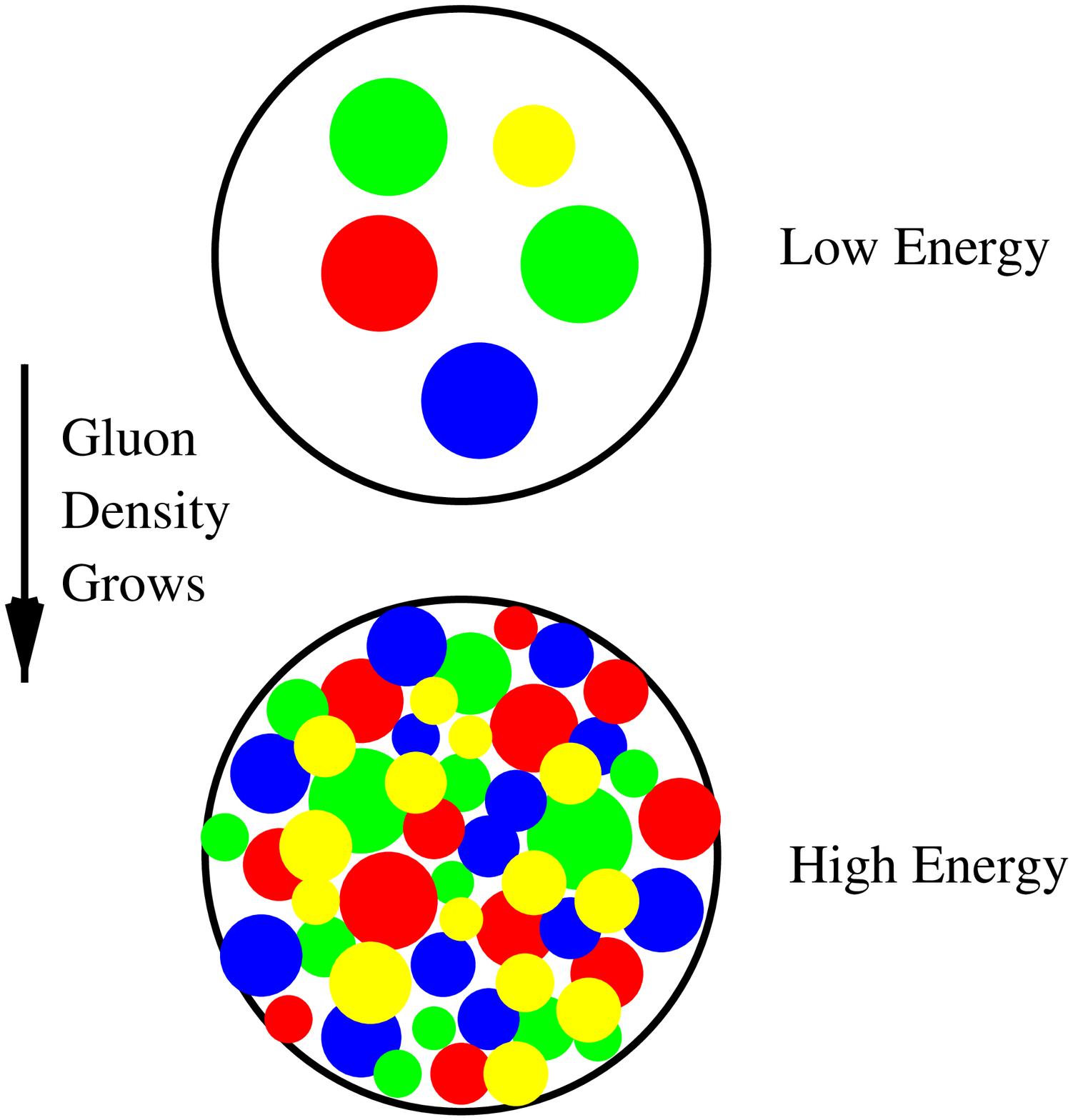,
        width=0.40\textwidth}}}
        \caption{(a)  The gluon structure function as a function
of x for various $Q^2$. (b)The increase in density of gluons as x 
decreases.  }
        \label{saturation}
\end{figure}

The gluon structure function $xG(x,Q^2)$ is experimentally measured
to increase at small x.  In the reference frame where the hadron is very
fast, $x$ is the ratio of a constituent energy to the projectile energy.
The gluon distribution is shown in  Fig. \ref{saturation}a.
Note the rapid increase in $xG(x,Q^2)$ as a function of $x$ for small x.
This is the origin of the ``small x problem''.  
This means that the piece of the hadron wavefunction relevant for 
small x processes has an increasing density of gluons.  In Fig. 
\ref{saturation}b, we look at a hadron headed along the beam direction.
As x decreases, the density of gluons increases.

The phase space density of gluons is 
\be
	\rho = {1 \over {\pi R^2}}  {{dN} \over {dyd^2p_T}}
\ee
where $R$ is the hadron size, $p_T$ is the transverse momentum of
a constituent, and $y \sim ln(1/x)$.  The high density of
gluons is generated dynamically and is caused by an instability, which is
proportional to the density.  The instability is stabilized when the density
of partons becomes large enough so that interactions of order 
$\alpha_{QCD} \eta ^2$
become of the order of the linear instability.  Here 
$\eta = \int~d^2p_T \rho$.  This requires that 
\be
	\eta \sim Q^2_{sat}/\alpha_{QCD}
\ee
The factor of $Q^2$ arises because we consider densities per unit area, and
$Q_{sat}^2$ carries this dimension.  This $Q_{sat}$ is called the saturation 
momentum.  The factor of $\alpha_{QCD}$ is the strong coupling strength of 
QCD.  When $Q_{sat} >> \Lambda_{QCD}$, we expect that 
$\alpha_{QCD} << 1$, so that the system becomes a high density Bose Condensate.

The name Color Glass Condensate arises therefore because
\begin{itemize}
\item{ \bf Color}

The gluons are colored.

\item{\bf Glass}
The natural time scale for the evolution of the gluon field is Lorentz 
time dilated.  This is like a glass which is a liquid on long times scales but
a solid on short ones.

\item{\bf Condensate}

The phase space density is as large as it can be.

\end{itemize}

\subsection{Space Time Evolution of Heavy Ion Collisions}

\begin{figure}[htb]
    \begin{center}
        \includegraphics[width=0.60\textwidth]{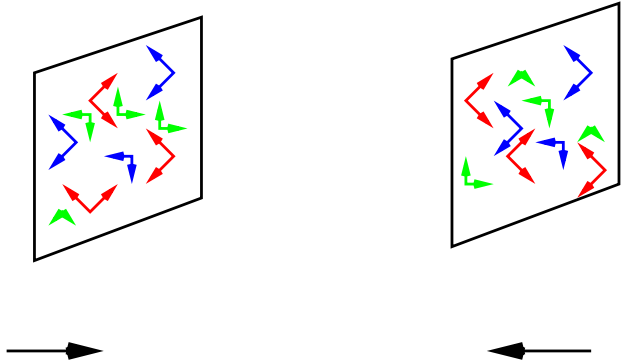}
        \caption{High energy nucleus-nucleus collisions.  
                            }\label{sheetonsheet}
    \end{center}
\end{figure}
A collision of two sheets of Colored Glass is shown in Fig. \ref{sheetonsheet}.

This is the picture of nucleus-nucleus collision which arises from the 
Color Glass Condensate.\cite{aa}  

The time evolution of the matter produced in these collisions is divided
into several stages:
\begin{itemize}

\item{\bf Initial Conditions}

For $t < 0$, the two sheets approach one another.  The Color Glass is frozen 
in each nucleus.  

\item{\bf Melting the Color Glass}

During the time $0 < t < t_{form}$, the Color Glass melts
into quarks and gluons.  It is estimated that $t_{form} \sim 
1/Q_{sat} \sim .1 - .3~ Fm/c$ at RHIC energy.  The energy density of the matter
at formation is somewhere around $\epsilon_{form} \sim Q^4_{sat} /\alpha_s
\sim 20 - 100~ GeV/Fm^3$.  

\item{\bf Thermalization}

During the time $t_{form} < t < t_{therm}$,
the matter expands and thermalizes.  Typical thermalization time is 
estimated to be $t_{therm} \sim .5 - 1 ~Fm/c$.  

\item{\bf Hydrodynamic Expansion}

The system expands 
as a thermal system until a time of decoupling which is typically about
$t_{decoupling} \sim 10 ~Fm/c$ at RHIC energy.  Here the matter
presumably starts as a Quark Gluon Plasma, evolves through a mixed
phase of hadrons and Quark Gluon Plasma and eventually becomes a
gas of pions.  In this stage, most of the physics interesting
for studies of the phase transition or cross over between
Quark Gluon Plasma and ordinary hadronic matter takes place.

\end{itemize}

\begin{figure}[htb]
    \begin{center}
        \includegraphics[width=0.40\textwidth]{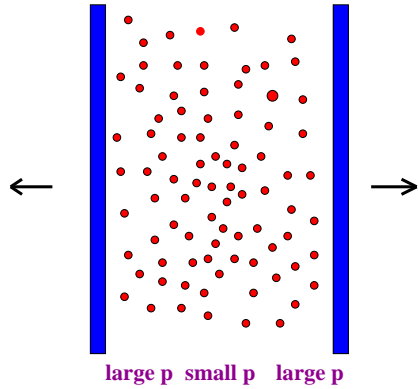}
        \caption{Particle production in nucleus-nucleus collision 
in the center of mass frame.}\label{collision}
    \end{center}
\end{figure}

As the Color Glass melts, it produces particles as is shown
in Fig. \ref{collision}.
The fastest particles have their natural time scale time dilated the most,
so in the center of mass frame, the fastest particles are produced last.
These particles have traveled the longest distance from the collision point,
since their formation time is dilated the most.
Therefore the matter is formed with a correlation between momentum and 
position.  This is like Hubble flow in cosmology where if you look at stars,
the stars which are farthest away move away the fastest.  For
heavy ions as in cosmology, this description is frame independent.
Unlike in cosmology, the Hubble expansion for heavy ion
collisions is 1 dimensional.\cite{bjorken}  

The density
of particles falls as $N/V \sim 1/t$.  If the particles expand without
interaction, then the energy per particle is constant.  If the particles
thermalize, then $E/N \sim T$, and since $N/V \sim T^3$ for a massless
gas, the temperature falls as $T \sim t^{-1/3}$.  For a gas which is not
quite massless, the temperature falls somewhere in the range $T_o
> T > T_o (t_o/t)^{1/3}$, that is the temperature is bracketed
by the value corresponding to no interaction and to that of a massless 
relativistic gas.  This 1 dimensional expansion continues
until the system begins to feel the effects of finite size in the 
transverse direction, and then rapidly cools through three dimensional
expansion.  Very close to when three dimensional expansion begins, the
system decouples and particle free stream without 
further interaction to detectors.

\section{What Have We Learned?} 

\subsection{The Energy Density is Big}

\begin{figure}[htb]
    \begin{center}
        \includegraphics[width=0.60\textwidth]{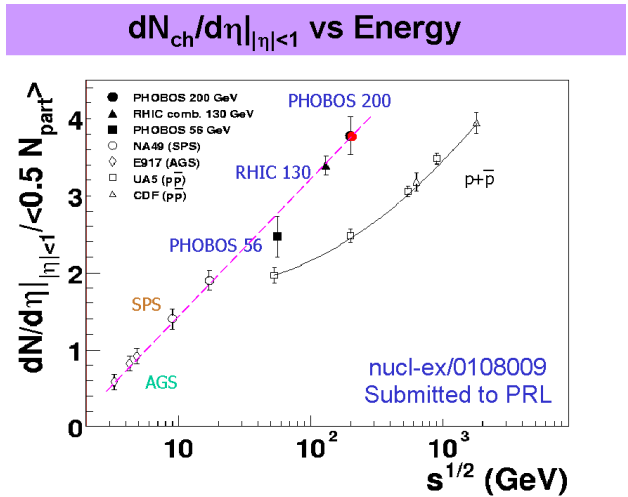}
        \caption{The multiplicity as a function of energy for
heavy ions and for $p \overline p$.  
                            }\label{dndye}
    \end{center}
\end{figure}

The  particle multiplicity as a function of energy has been measured at 
RHIC and is shown in Fig. \ref{dndye}.\cite{star}-\cite{brahms}  
On the same plot is shown lower energy data from the AGS and CERN
and data from the $p\overline p$ collider as well.  One sees that the energy
dependence in $AA$ collisions is different than that for $p \overline p$,
and is more or less consistent with a $ln(s)$ behavior.
\begin{figure}[htb]
    \begin{center}
        \includegraphics[width=0.80\textwidth]{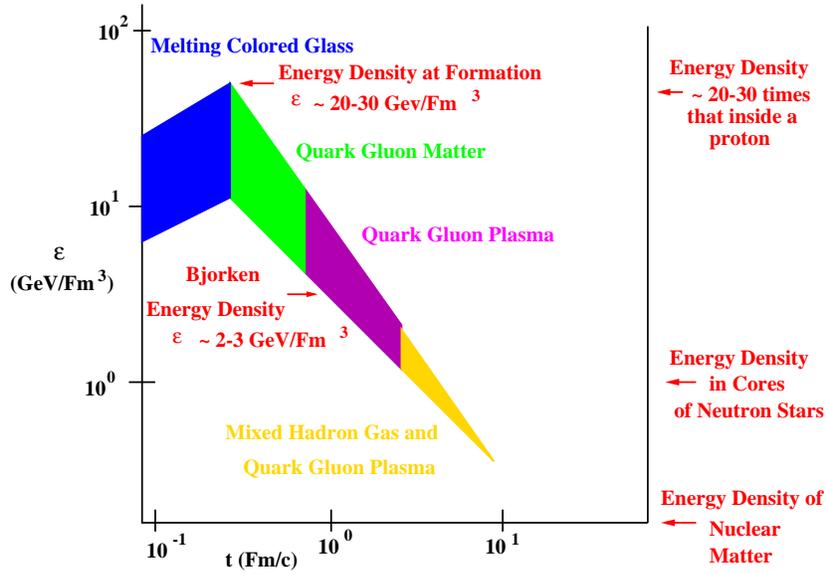}
        \caption{Bounds on the energy density as a function of time
in heavy ion collisions.  
                            }\label{bounds}   
 \end{center}
\end{figure}

Combining the multiplicity data together with the measurements of transverse
energy or of typical particle transverse momenta, one can determine the
energy density of the matter when it decouples.  One can then extrapolate
backwards in time.  We extrapolate backwards using 1 dimensional 
expansion, since decoupling occurs when the matter first begins
to expands three dimensionally.  We can extrapolate backwards until  
the matter has melted from a Color Glass.  We shall take a conservative 
overestimate of this time to be of order $t_{melt} \sim .3 ~Fm/c$  
The
extrapolation backwards is bounded by $\epsilon_{f} (t_f/t)
< \epsilon(t) < \epsilon_f (t_f/t)^{4/3}$.  The lower bound is
that assuming that the particles do not thermalize and their typical
energy is frozen.  The upper bound assumes that the system thermalizes
as an ideal massless gas.  We argued above that the true result is 
somewhere in between.  When the time is as small as the melting time,
then the energy density begins to decrease as time is further decreased.

This bound on the energy density is shown in Fig. \ref{bounds}.
On the left axis is the energy density and on the bottom axis is time.
The system begins as a Color Glass Condensate, then melts to Quark
Gluon Matter which eventually thermalizes to a Quark Gluon Plasma.
At a time of a few $Fm/c$, the plasma becomes a mixture of quarks, gluons and 
hadrons which expand together.  
\begin{figure}[htb]
    \begin{center}
        \includegraphics[width=0.80\textwidth]{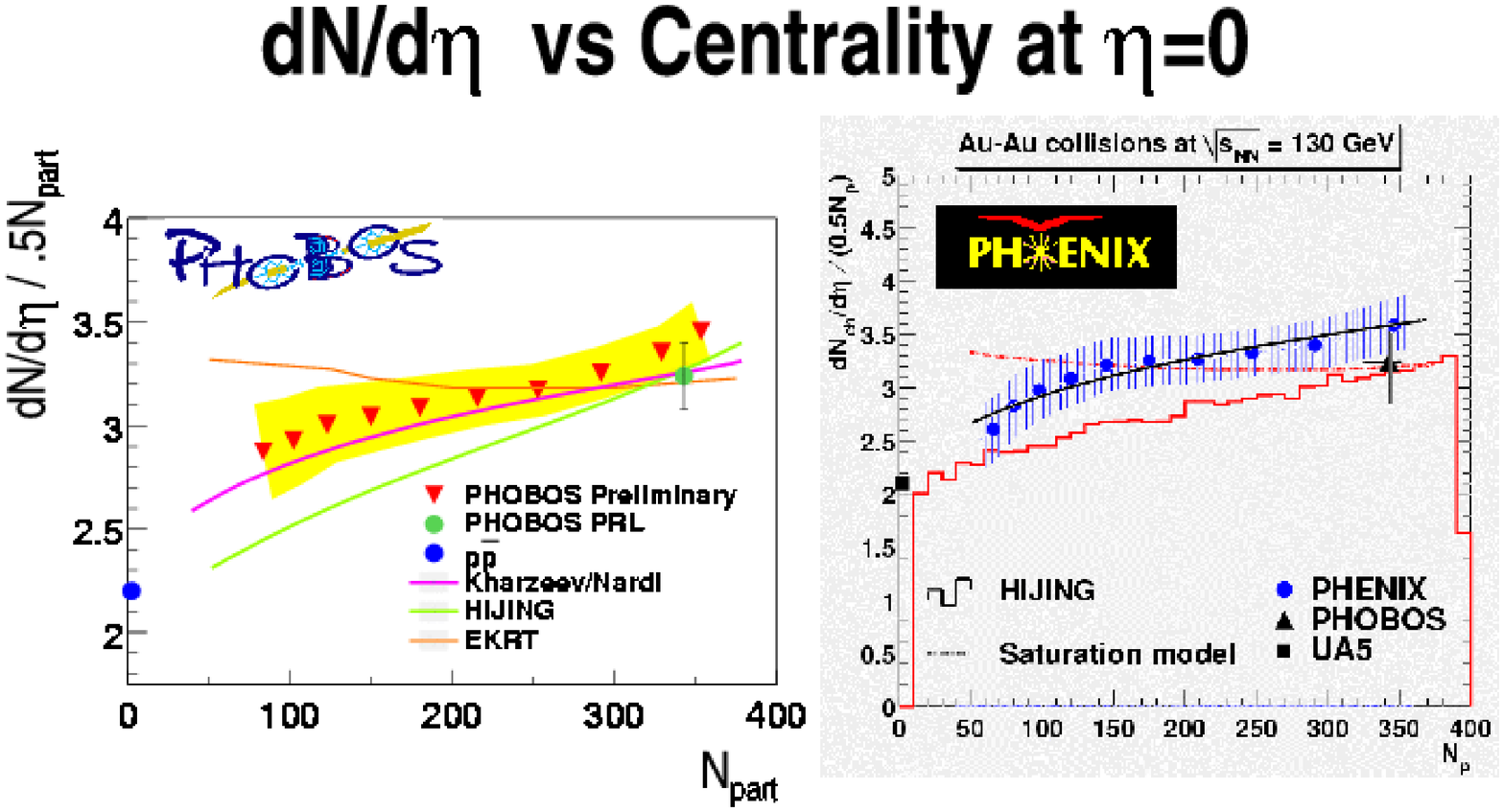}
        \caption{The multiplicity per participant as a function
of the number of participants.  The solid line running through the
center of both of the sets of data is the Kharzeev-Nardi result.  The
EKRT result is marked as such in the left hand plot and is called the
saturation model in the right hand plot.  HIJING is marked on both plots. 
                            }\label{npart}   
 \end{center}
\end{figure}
At a time of about $10 ~Fm/c$, the
system falls apart and decouples.  At a time of $t \sim 1~Fm/c$,
the estimate we make is identical to the Bjorken energy density estimate,
and this provides a lower bound on the energy density achieved 
in the collision.  (All estimates agree that by a time of order $1 ~Fm/c$,
matter has been formed.)  The upper bound corresponds to assuming
that the system expands as a massless thermal gas from a melting
time of $.3~Fm/c$.  (If the time was reduced, the upper bound would  
be increased yet further.)
The bounds on the energy density 
are therefore
\be
	2-3 ~GeV/Fm^3 < \epsilon < 20-100 ~GeV/Fm^3
\ee
where we included a greater range of uncertainty in the upper limit
because of the uncertainty associated with the formation time.
The energy density of nuclear matter is about $0.15~GeV/Fm^3$, and
even the lowest energy densities in these collisions is in excess of this.
At late times, the energy density is about that of the cores of neutron stars,
$\epsilon \sim 1 ~GeV/Fm^3$.

{\bf At such extremely high energy densities, 
it is silly to try do describe the
matter in terms of anything but its quark and gluon degrees of freedom.}

\subsection{The Gross Features of Multiplicity Distributions Are Consistent
with Colored Glass}
\begin{figure}[htb]
    \centering
       \mbox{{\epsfig{figure=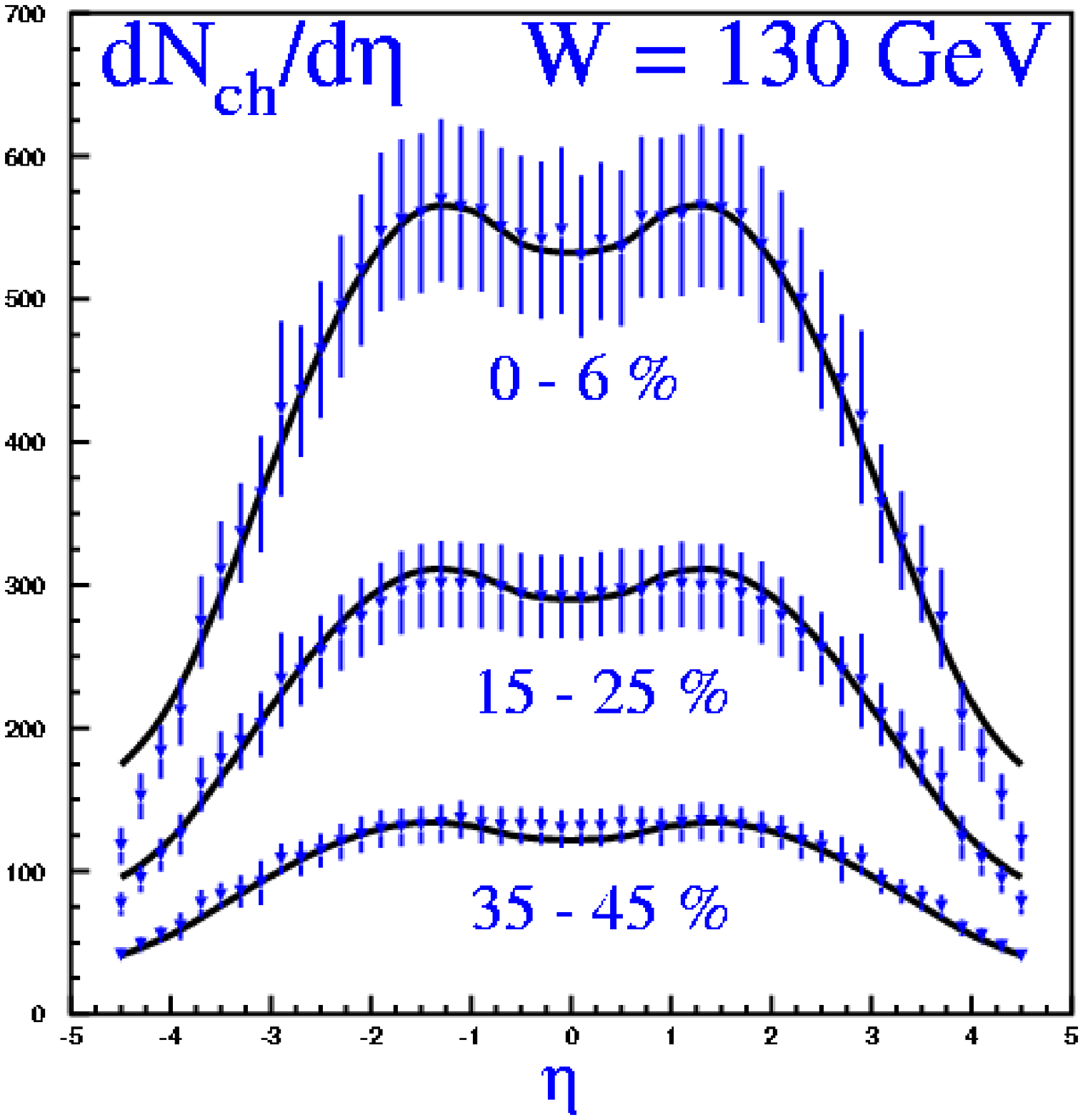,
        width=0.50\textwidth}}\quad
             {\epsfig{figure=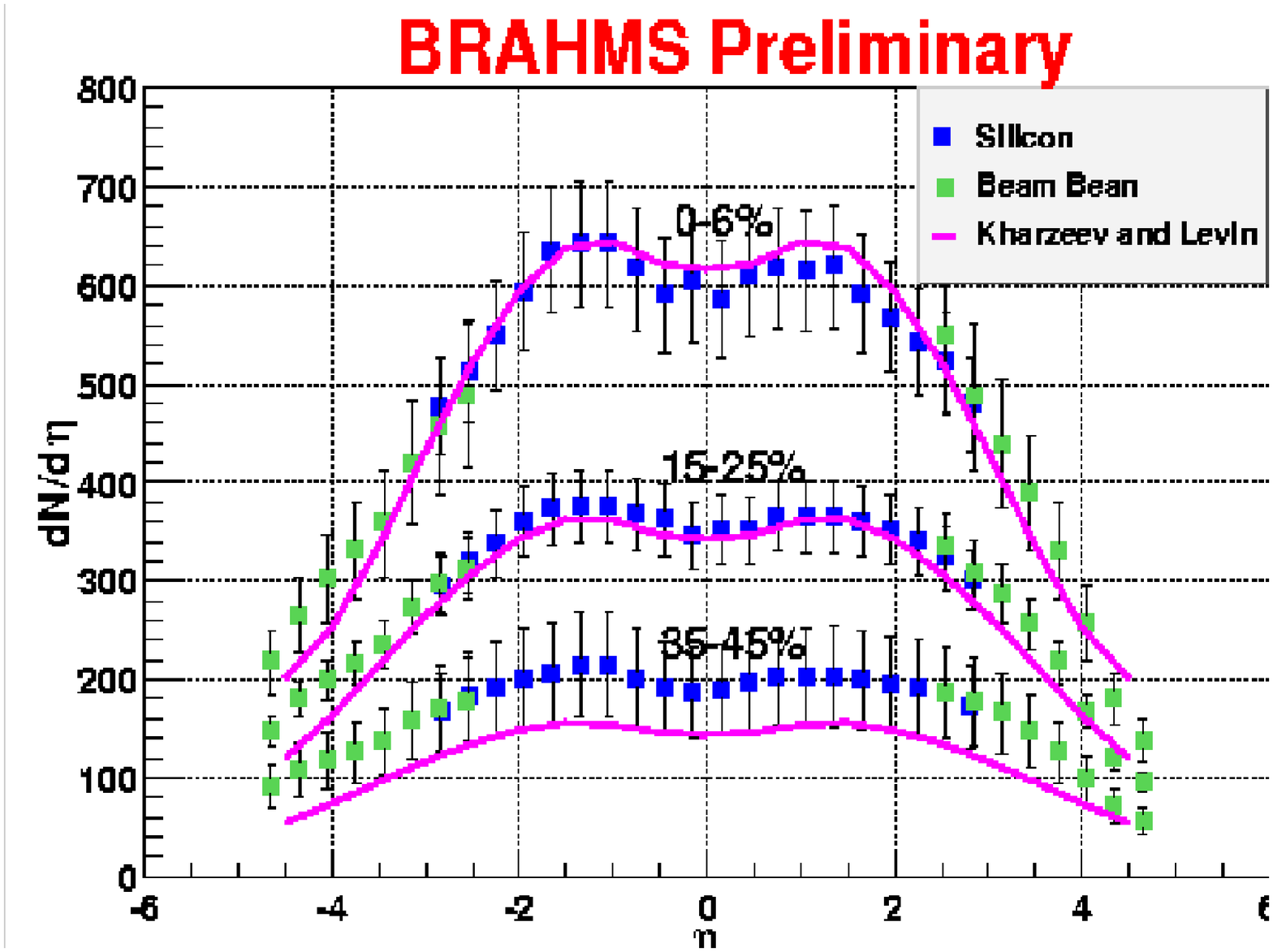,
        width=0.50\textwidth}}}
        \caption{Color glass condensate fits to the rapidity density
measured in the PHOBOS and Brahms experiments}
        \label{dndysat}
\end{figure}

As argued by Kharzeev and Nardi,\cite{kn} the density of produced
particles per nucleon which participates in the collision,
$N_{part}$,  should be proportional to
$1/\alpha_s(Q_{sat})$, and $Q_{sat}^2 \sim N_{part}$. This dependence 
follows from the $1/\alpha_s$ which characterizes the density of
the Color Glass Condensate.
In Fig. \ref{npart}, we show the 
data from PHENIX and PHOBOS\cite{npart}.  The 
Kharzeev-Nardi form fits the data well.  Other attempts such as 
HIJING\cite{hijing},
or the so called saturation model of 
Eskola-Kajantie-Ruuskanen-Tuominen\cite{ekrt} are also shown in the
figure.

Kharzeev and Levin have recently argued that the rapidity distributions
as a function of centrality can be computed from the 
Color Glass description.\cite{kl}  This is shown in 
Fig. \ref{dndysat}.\cite{phbr}

\subsection{Matter Has Been Produced which Interacts Strongly 
with Itself}

\begin{figure}[htb]
    \begin{center}
        \includegraphics[width=0.80\textwidth]{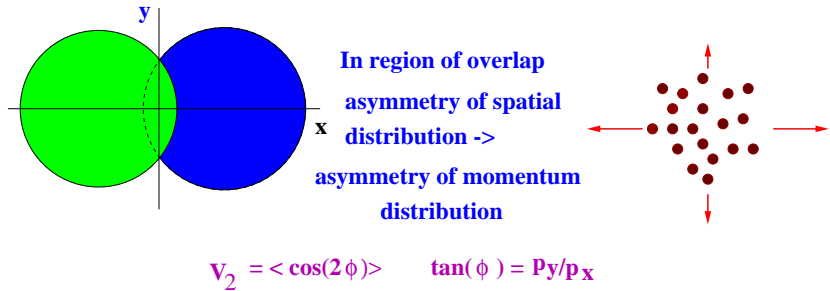}
        \caption{The asymmetry in the distribution of matter in an off center
collision is converted to an asymmetry of the momentum space distribution.  
                            }\label{flow}   
 \end{center}
\end{figure}

\begin{figure}[htb]
    \centering
       \mbox{{\epsfig{figure=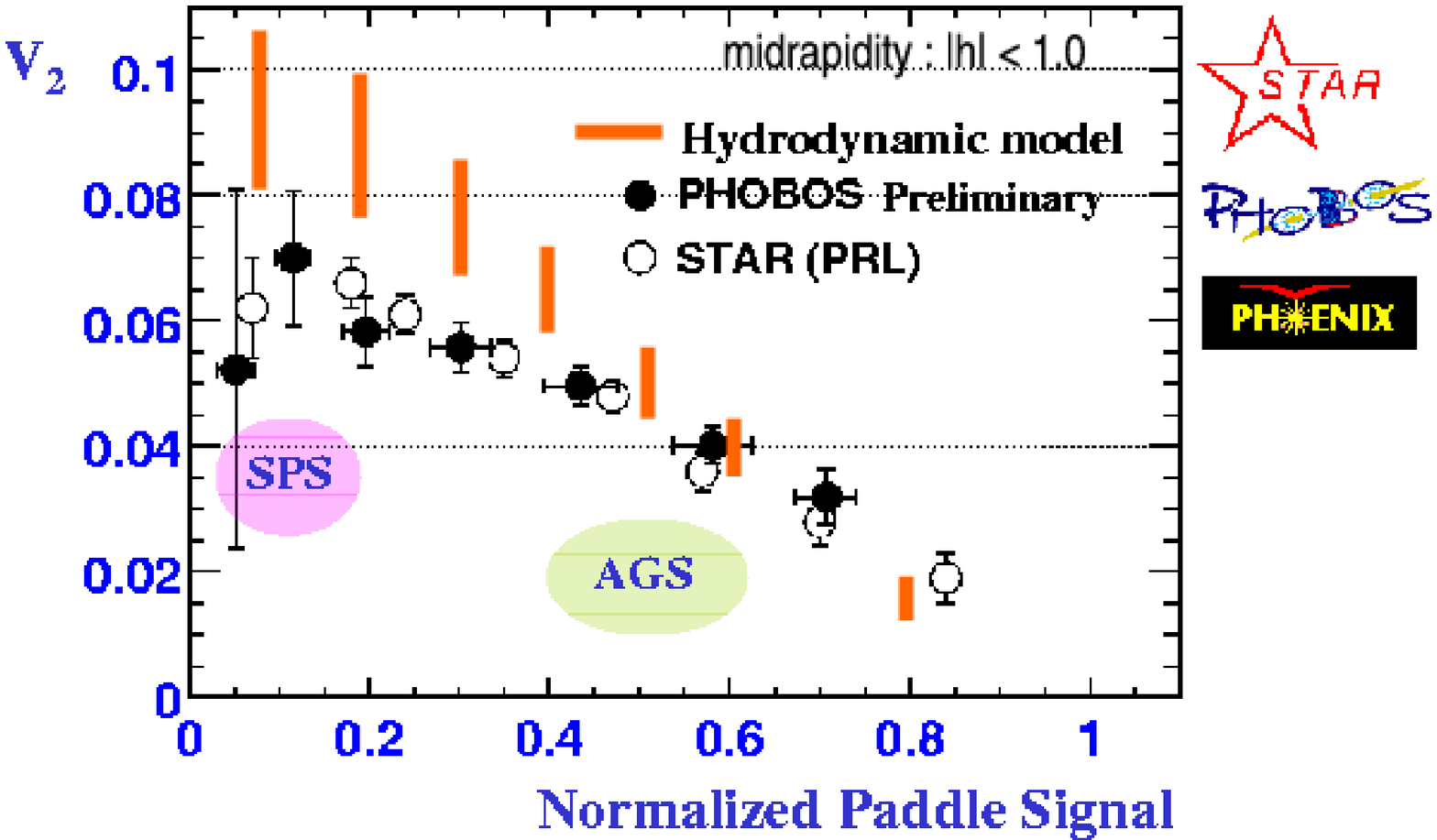,
        width=0.50\textwidth}}\quad
             {\epsfig{figure=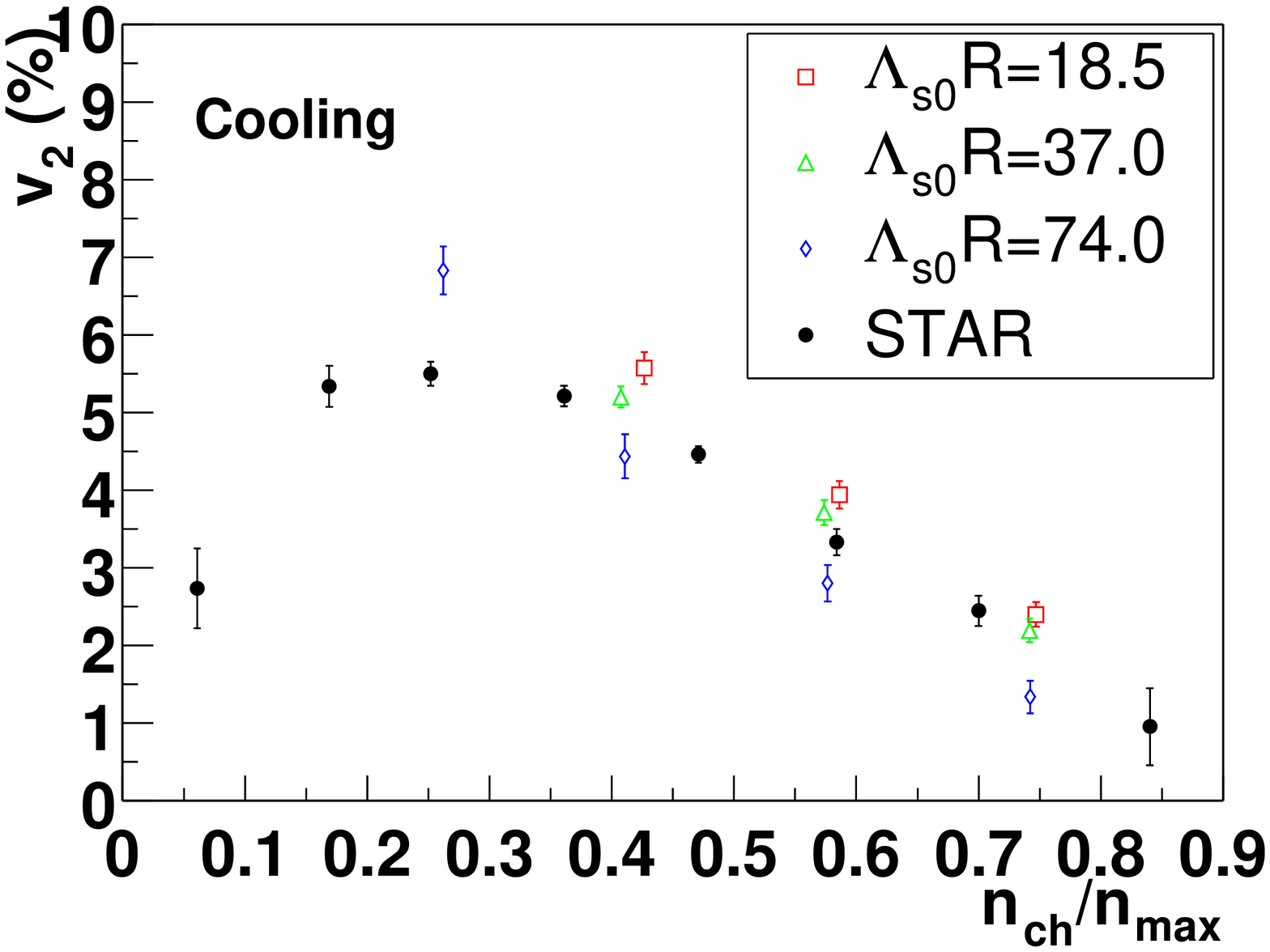,
        width=0.50\textwidth}}}
        \caption{a) A hydrodynamic fit to v2.  b) The Colored Glass fit.}
        \label{flow1}
\end{figure}

In off zero impact parameter heavy ion collisions, the matter which
overlaps has an asymmetry in density relative to the reaction plane.
This is shown in the left hand side of Fig. \ref{flow}.  Here
the reaction plane is along the x axis.  In the region of overlap there
is an $x-y$ asymmetry in the density of matter which overlaps.
This means that there will be an asymmetry in the spatial gradients which
will eventually transmute itself into an asymmetry in the momentum
space distribution of particles, as shown in the right hand side of
Fig. \ref{flow}.
This asymmetry is called elliptic
flow and is quantified by the variable $v_2$.  In all
attempts to theoretically describe this effect, one needs very strong
interactions among the quarks and gluons at very early times in the 
collision.\cite{flowth}.  In Fig. \ref{flow1}, two different theoretical 
descriptions are fit to the data by STAR and 
PHOBOS\cite{flowstar}-\cite{flowphobos}. On the left hand side, a 
hydrodynamical model is used.\cite{heinzkolb}  
It is roughly of the correct order of magnitude
and has roughly the correct shape to fit the data.  This was not
the case at lower energy.  On the right hand side are preliminary
fits assuming Color 
Glass.\cite{rajualex}  
Again it has roughly the correct shape and magnitude to describe
the data.  In the Color Glass, the interactions are very strong essentially
from $t = 0$, but unlike the hydrodynamic models it is field pressure
rather than particle pressure which converts the spatial anisotropy into
a momentum space-anisotropy.  

Probably the correct story for describing flow
is complicated and will involve both the Quark Gluon Plasma and the
Color Glass Condensate.  Either description requires that matter be produced
in the collisions and that it interacts strongly with itself.  In the limit
of single scatterings for the partons in the two nuclei, no flow is generated.

\section{What Do We Expect to Learn?}

\subsection{Does the Matter Equilibrate?}

\begin{figure}[htb]
    \centering
       \mbox{{\epsfig{figure=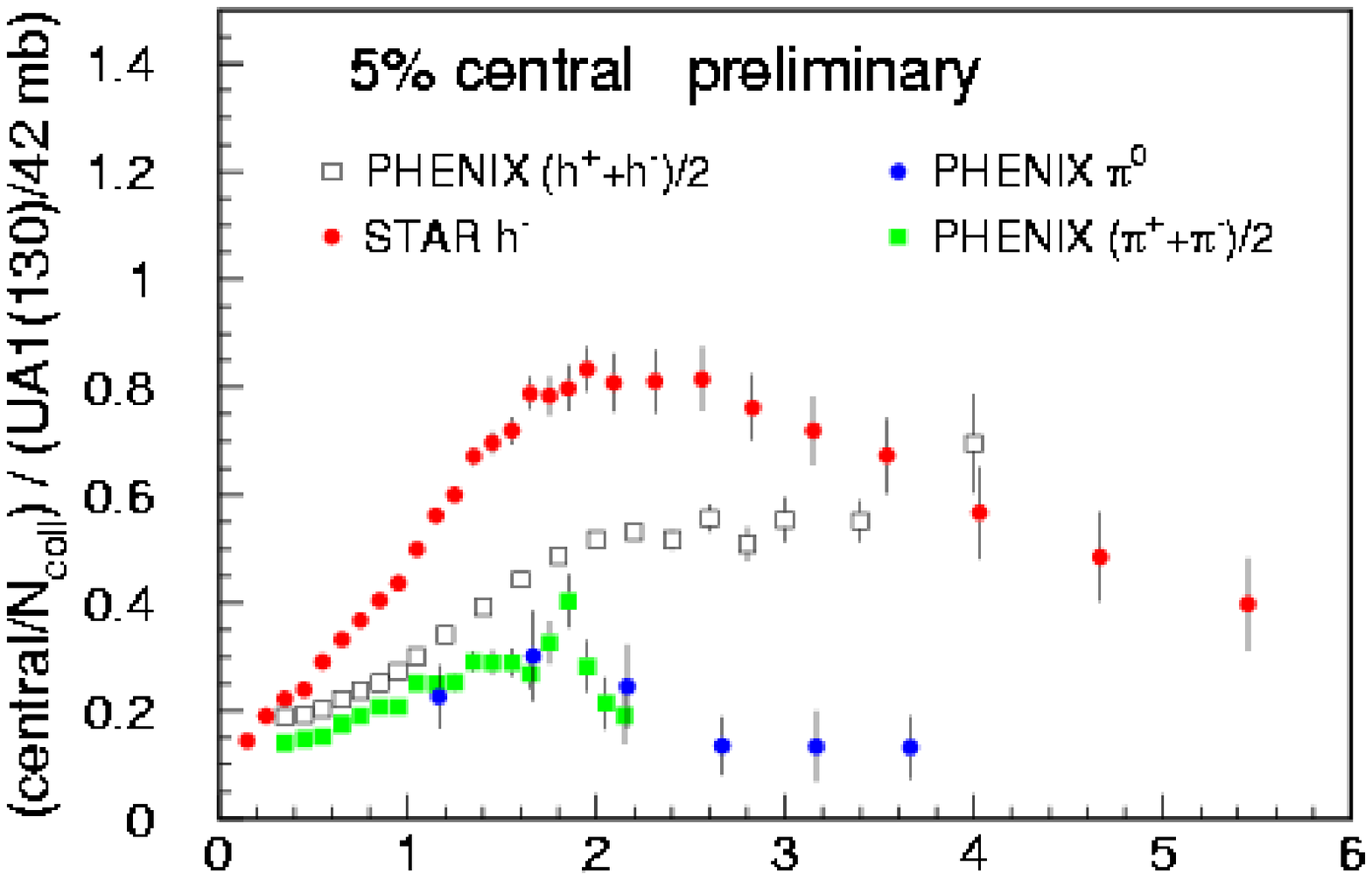,
        width=0.50\textwidth}}\quad
             {\epsfig{figure=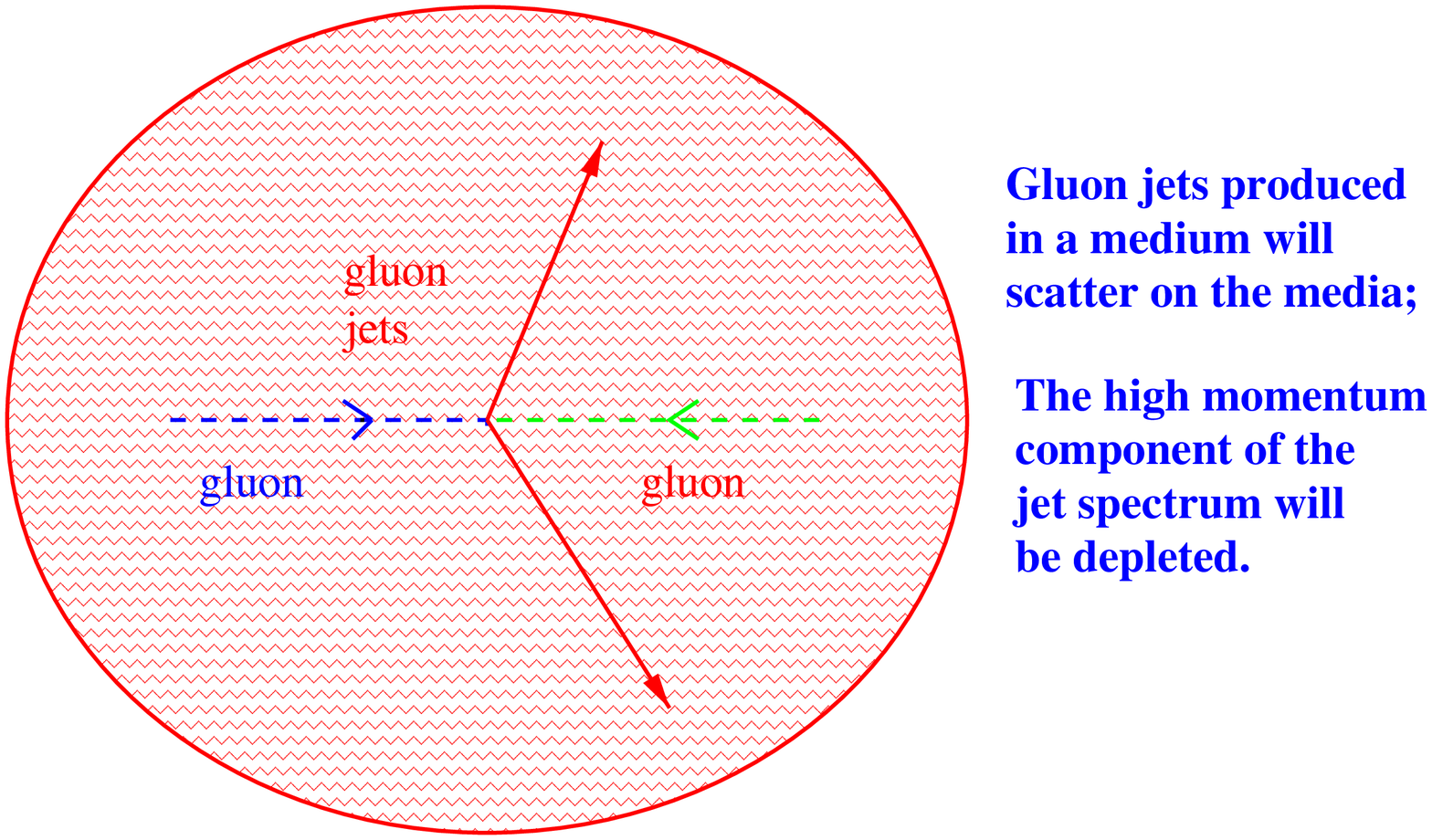,
        width=0.50\textwidth}}}
        \caption{a) The $p_T$ distribution of particles
scaled by the data from $pp$ collisions times the number of hard collisions
inside the nuclei.  b) A pair of jets is produced in a 
hard collision and they propagate through the surrounding matter.}
        \label{jets}
\end{figure}

One of the most interesting results from the RHIC experiments
is the so called ``jet quenching''.\cite{starjet}-\cite{phenixjet}.
In Fig. \ref{jets}a, the single particle hadron spectrum is scaled
by the same result in $pp$ collisions and scaled by the number of
collisions.  The number of collisions is the number of nucleon-nucleon 
interactions, which for central collisions should be almost all of the 
nucleons.  One is assuming hard scattering in computing this number,
so that a single nucleon can hard scatter a number of times as it penetrates 
the other nucleus.  The striking feature of this plot is that
the ratio does not approach one at large $p_T$.  This would be the value
if these particles arose from hard scattering which produced jets
of quarks and gluons, and the jets subsequently decayed.

The popular explanation for this is shown in Fig. \ref{jets}b.  Here a
pair of jets is produced in a gluon-gluon collision.  The jets are
immersed in a Quark Gluon Plasma, and rescatter as they poke through
the plasma.  This shifts the transverse momentum spectrum down, and the
ratio to $pp$ collisions, where there is no significant surrounding media,
is reduced.

The data, however suggestive, need to be improved before strong 
conclusions are drawn.  For example, there are large systematic
uncertainties in the $pp$ data which was measured in different detectors
and extrapolated to RHIC energy.  This will be resolved by measuring $pp$ 
collisions at RHIC.  There is in addition 
some significant uncertainty in the AA data which becomes smaller
in the ratio to $pp$ data when the data is measured in the same detector.
There are nuclear modifications of the gluon distribution function,
an effect which can be determined by measurements on $pA$ at RHIC.
The maximum transverse momentum is limited by the event sample size,
and the size will be greatly improved with this years run due to the
higher luminosity and longer run time.

One of the reasons why jet quenching is so important for the RHIC
program is that it gives a good measure of the degree of thermalization
in the collisions.  If jets are strongly quenched by transverse momenta of
$4 ~GeV$, then because cross sections go like $1/E^2$ for quarks and gluons,
this would be strong evidence for thermalization at the lower energies
typical of the emitted particles.

One can look for evidence of thermalization directly from the measured $p_T$
distributions.  Here one can do a hydrodynamic computation and in so far as
it agrees with the results, one is encouraged to believe that there is
thermalization.  On the other hand, these distributions may have their
origin in the initial conditions for the collision, the Colored Glass.
In reality, one will have to understand the tradeoff between both effects.

\begin{figure}[htb]
    \centering
       \mbox{{\epsfig{figure=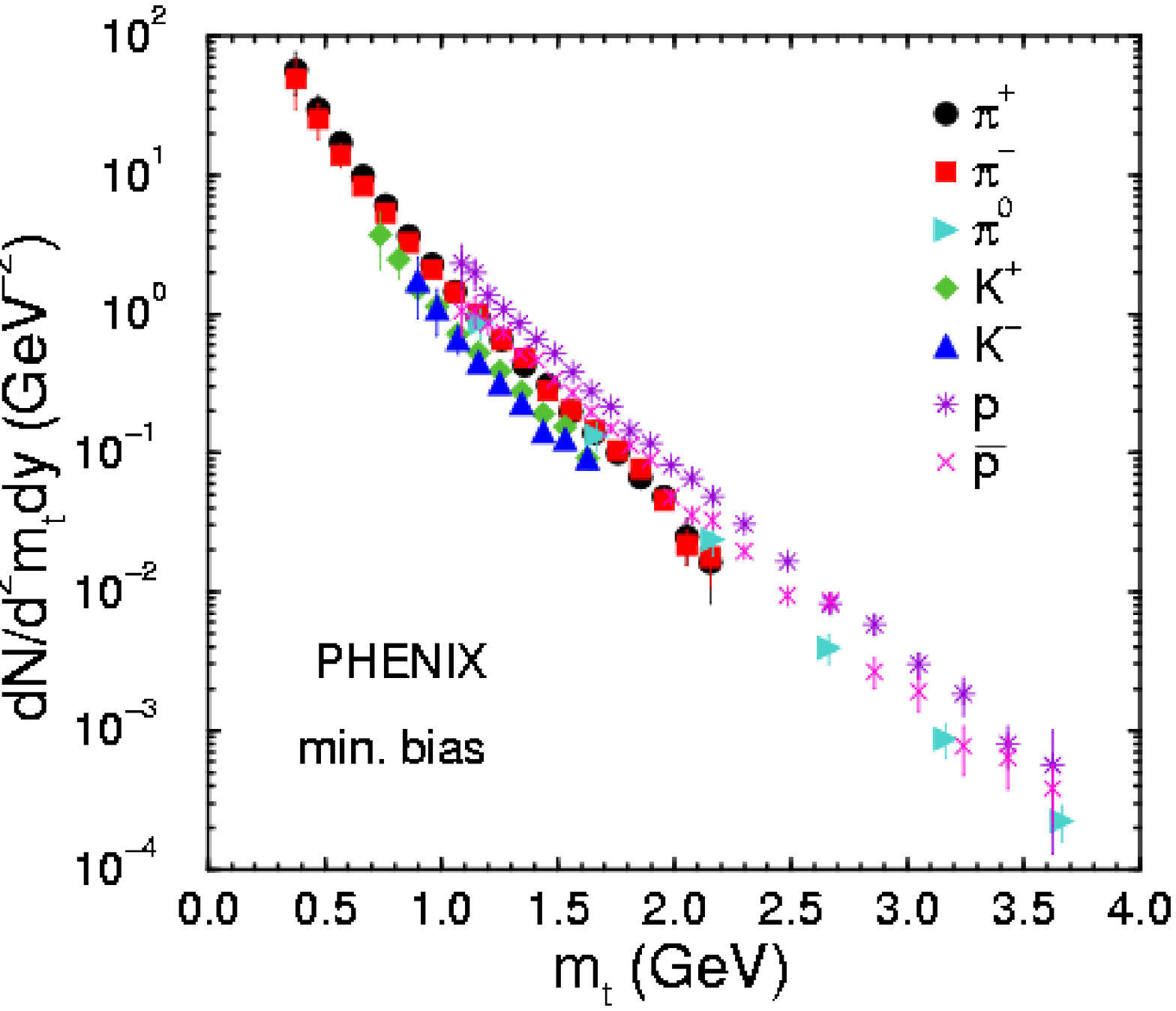,
        width=0.50\textwidth}}\quad
             {\epsfig{figure=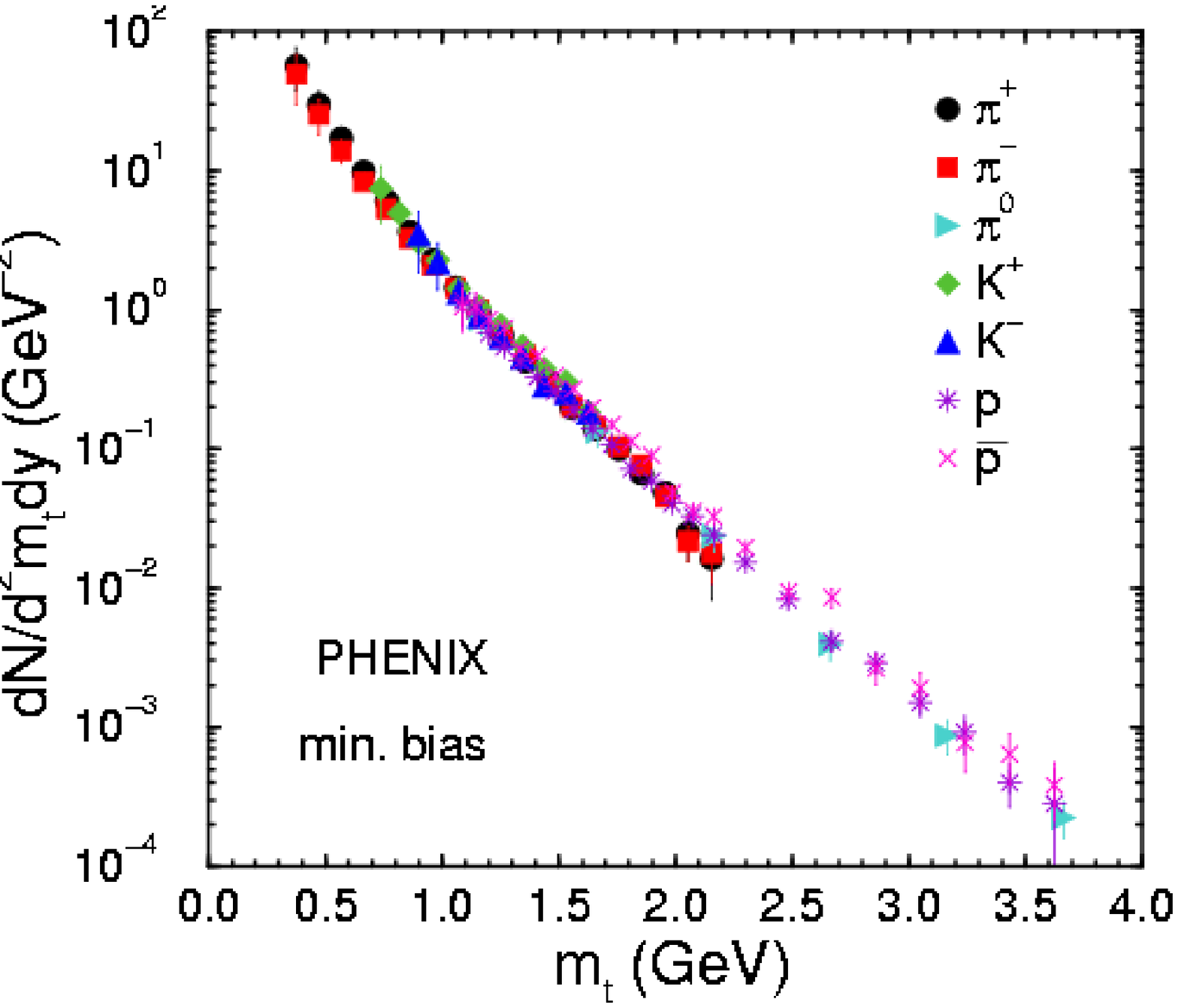,
        width=0.50\textwidth}}}
        \caption{a The $m_T$ distributions of identified particles.
b) The distributions rescaled by a constant dependent 
upon particle species.}
        \label{mt}
\end{figure}

Let us begin with the Color Glass description.  In Fig. \ref{mt}a,
the $mT$ distributions ($m_T^2 = p_T^2 + m^2$)
of identified particle measured for minimum bias events 
in PHENIX are shown.\cite{phenixjet}
In Fig \ref{mt}b, we replot these curves by rescaling by a constant
multiplicative factor for each of the particles.  
In thermal models, this constant factor would be associated with 
chemical potentials for producing different particle species.
These plots fall on top
of one another.  The distributions therefore scale in $m_T$.

\begin{figure}[htb]
    \centering
       \mbox{{\epsfig{figure=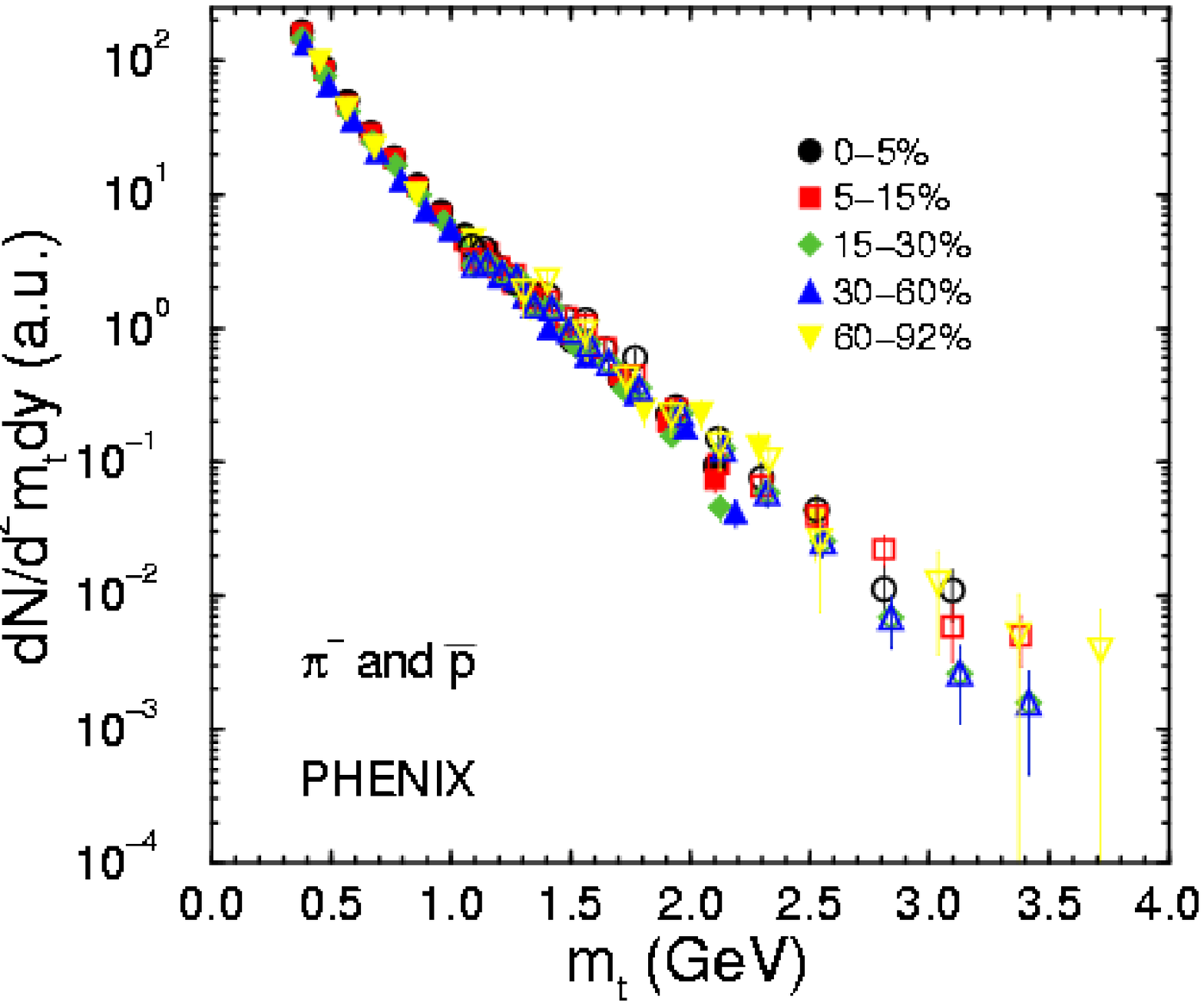,
        width=0.50\textwidth}}\quad
             {\epsfig{figure=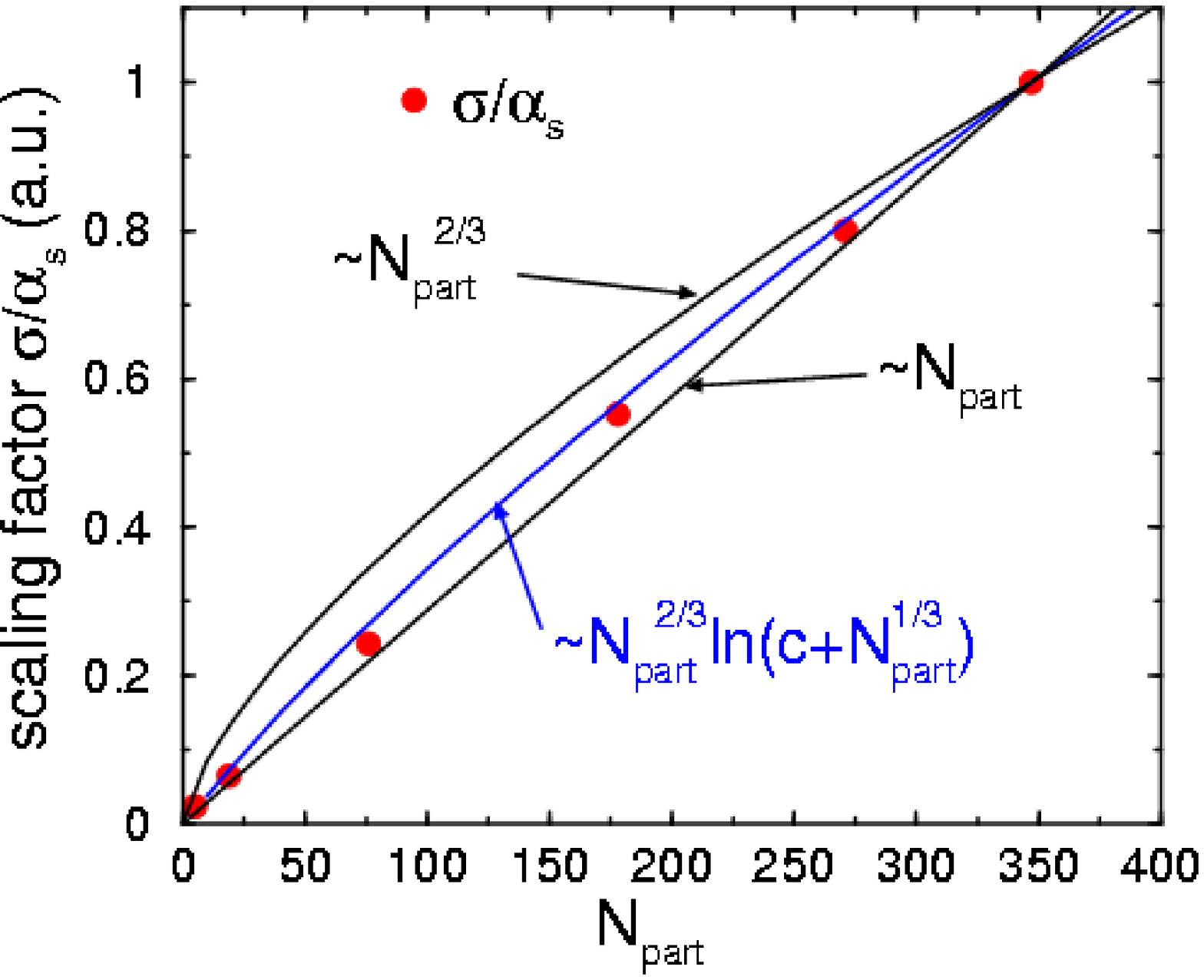,
        width=0.50\textwidth}}}
        \caption{a) The $m_T$ distributions for various centralities
have been rescaled into one another.  b) The dependence of the
coupling $1/\alpha_s$ is compared to the empirically determined scaling factor
for the $m_T$ distributions.}
        \label{mt1}
\end{figure}

If the distributions scale in $m_T$, as they appear to do for $m_T$ within
the range measured by PHENIX, then the Colored Glass predicts as scaling
relation for the distributions:\cite{schaffner}
\be
	{{dN} \over {dyd^2p_T}} = {{\pi R^2} \over \alpha_s } F (m_T/Q_{sat})
\ee
This means that if we  adjust  the normalization of the distribution
function and the scale of its dependence on $m_T$ 
then the distributions should
fall on top of one another.  The scale adjustment should be proportional
to $1/\alpha_s$ which should in turn involve $Q_{sat}$ from the running of
the coupling.  In Fig. \ref{mt1}a, the data at different centralities
are rescaled into one another according to the above relation.  This seems
to describe the data well.  In Fig. \ref{mt1}b, the dependence of
$1/\alpha_s$ extracted from the above equation is compared to what is
determined from the running coupling constant.  Again, this works quite well.
The PHENIX data is consistent with the scaling relations predicted by the
Color Glass Condensate. 

\begin{figure}[htb]
    \begin{center}
        \includegraphics[width=0.50\textwidth]{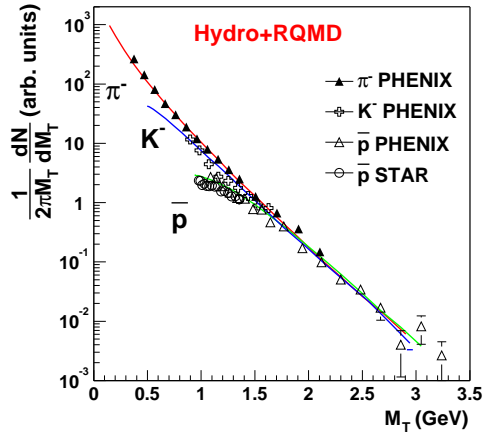}
        \caption{The hydrodynamical model fits to the $m_T$ spectra
for the PHENIX and STAR data.  
                            }\label{teaney}   
 \end{center}
\end{figure}

Hydrodynamical models also successfully describe the data on 
$m_T$\linebreak  distributions.\cite{shuryak}  In Fig. \ref{teaney}
the results of the simulation by Shuryak and Teaney are shown compared
to the STAR and PHENIX data.\cite{starjet}-\cite{phenixjet}  The shape of the
curve is a prediction of the hydrodynamic model, and is parameterized somewhat
by the nature of the equation of state.  Notice that the
STAR data include protons near threshold, and here the $m_T$ scaling 
breaks down.  The hydrodynamic fits get this dependence correctly,
and this is one of the best tests of this description.
The hydrodynamic models do less well on fits to the more peripheral collisions,
but the Colored Glass model gets this more or less correctly.  In general,
a good place to test the hydrodynamic models predictions is with massive
particles close to threshold, since here one deviates in a computable way
from the $m_T$ scaling curve, which is arguably determined from 
parameterizing the equation of state.

If one can successfully argue that there is thermalization in the RHIC
collisions, then the hydrodynamic computations become compelling.
One should remember that hydrodynamics requires an equation of state plus 
initial conditions, and these initial conditions are determined by Colored
Glass.  Presumably, a correct description will require the inclusion of both
types of effects.\cite{srivastava}

\section{What Do We Hope to Learn?}

\subsection{Confinement and Chiral Symmetry Restoration}
We would like to know whether or not deconfinement has occurred in dense matter
or whether chiral symmetry restoration has changed particle masses.
\begin{figure}[htb]
    \begin{center}
        \epsfxsize = 3in
        \epsfbox[20 100 550 700]{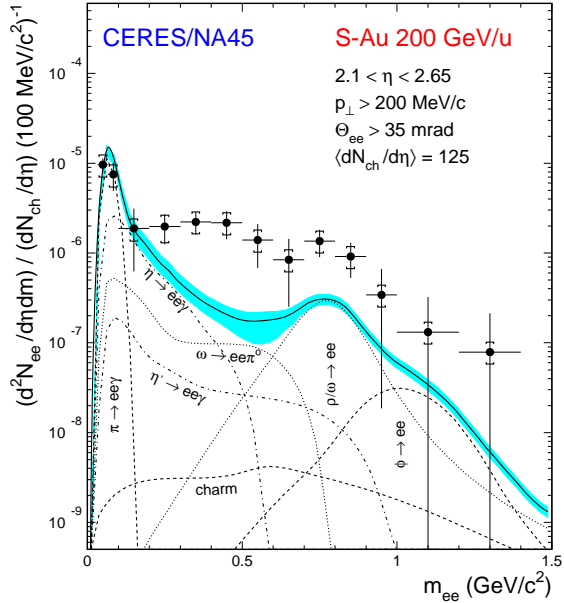}
        \caption{The CERES data on low mass electron-positron pairs.  
The expected 
contribution from ordinary hadrons is shown by the solid line. The data 
points are for the measured electron-positron pairs.   
                            }\label{ceres1}   
 \end{center}
\end{figure}


This can be studied in principle by measuring the spectrum of dileptons
emitted from the heavy ion collision.  These particles probe the interior
of the hot matter since electromagnetically interacting particles are
not significantly attenuated by the hadronic matter.  For electron-positron 
pairs, the mass distribution has been measured in the CERES experiment at 
CERN\cite{ceres}, and is shown in Fig. \ref{ceres1}.  Shown in the plot
is the distribution predicted from extrapolating from $pA$ collisions.
There should be a clear $\rho $ and $\phi $ peak, which has disappeared.
This has been interpreted as a resonance mass shift,\cite{brown},
enhanced $\eta^\prime$ production, \cite{kapusta} but is most probably
collisional broadening of the resonances in the matter
produced in the collisions.\cite{rapp}  Nevertheless, if one makes a plot
such as this and the energy density is very high and there are no resonances
at all, then this would be strong evidence that the matter is not hadronic, 
i. e. the hadrons have melted.  

The resolution in the CERES experiment is unpleasantly large, making it
difficult to unambiguously interpret the result.  Whether or not
such an experiment could be successfully run at RHIC, much less whether
the resolution could be improved, is the subject of much internal debate 
among the RHIC experimentalists.  

\subsection{Confinement and $J/\Psi$ Suppression}

\begin{figure}[htb]
    \centering
       \mbox{{\epsfig{figure=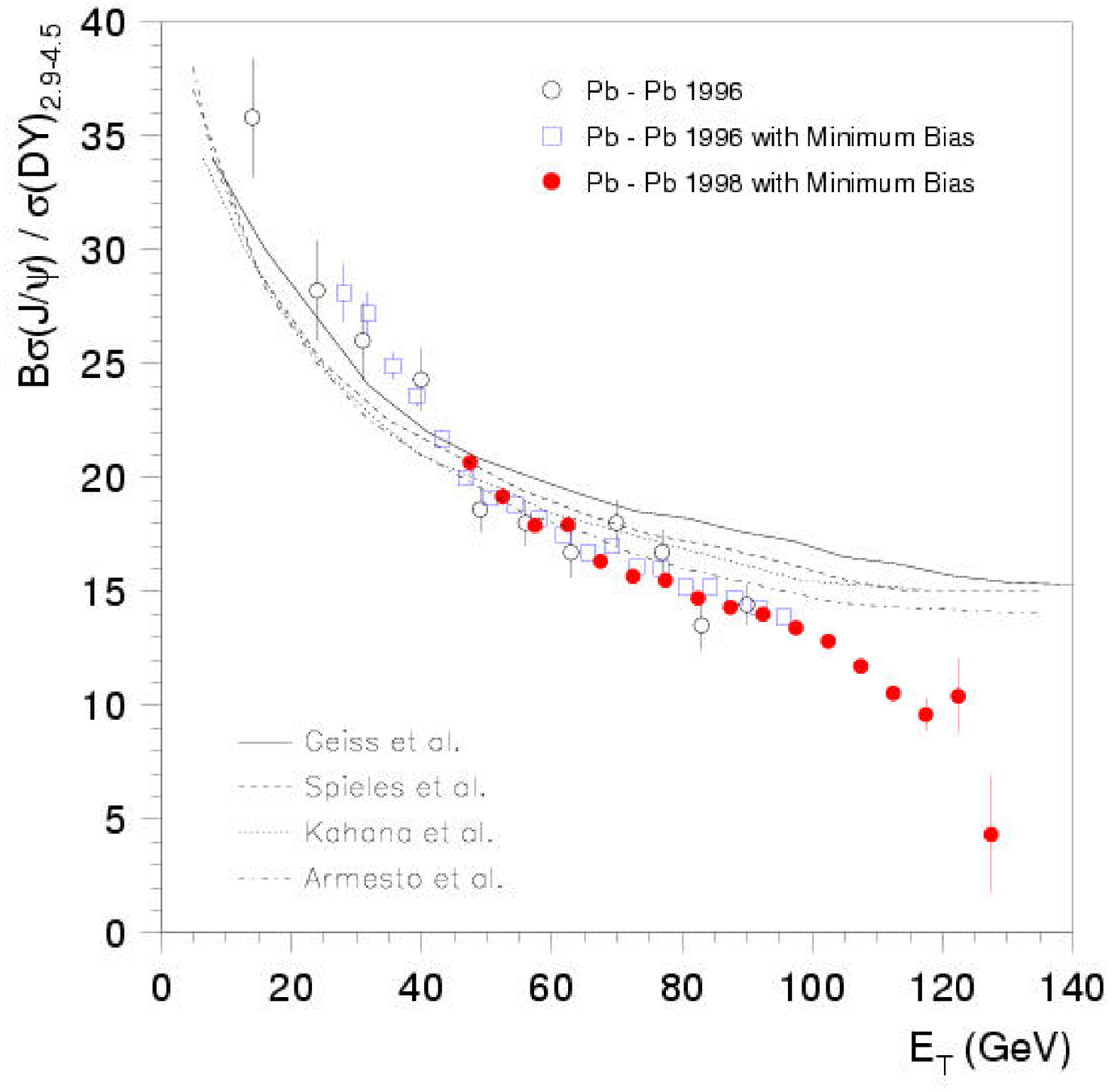,
        width=0.50\textwidth}}\quad
             {\epsfig{figure=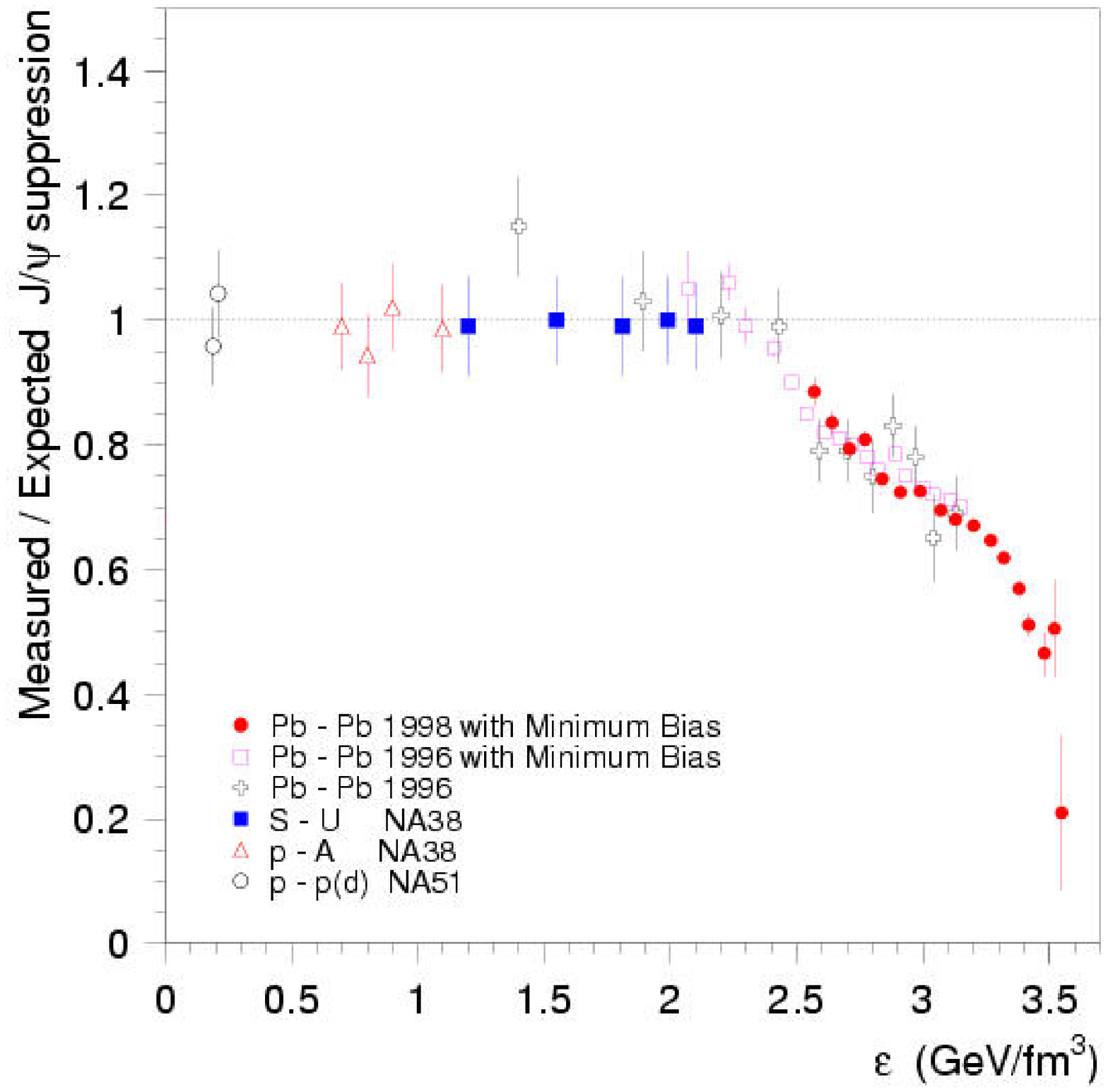,
        width=0.50\textwidth}}}
        \caption{a) 
The ratio of produced $J/\Psi$ pairs to Drell-Yan pairs as a function of 
transverse energy $E_T$ at CERN energy.  b) The measured
compared to the theoretically expected $J/\Psi$ suppression as a function
of the Bjorken energy density for various targets and projectiles.}
        \label{jpsi}
\end{figure}

In Fig. \ref{jpsi},  the NA(50) data for $J/\Psi$ production is 
shown.\cite{na50}
In the first figure, the ratio of $J/\Psi$ production cross section
to that of Drell-Yan is shown as a function of $E_T$,
the transverse energy, for the lead-lead 
collisions at CERN.  There is a clear suppression at large $E_T$ which
is greater than the hadronic absorption model calculations which are plotted
with the data.\cite{hadab}  In the next figure, the 
theoretically expected $J/\Psi$ suppression 
based on hadronic absorption models is compared
to that measured as a function of the Bjorken energy density
for various targets and projectiles.  There appears to be a sharp increase
in the amount of suppression for central lead-lead collisions.

Is this suppression due to Debye screening of the confinement potential
in a high density Quark Gluon Plasma?\cite{matsui}-\cite{blaizot}
Might it be due higher twists, enhanced rescattering, or changes in the gluon
distribution function?\cite{capella}-\cite{qiu}  Might the $J/\psi$
suppression be changed into an enhancement 
at RHIC energies and result from the 
recombination in the produced charm particles, many more of which are
produced at RHIC energy?\cite{rafelski}-\cite{gorenstein}

These various descriptions can be tested by using the    
capability at RHIC to do $pp$ and $pA$ as well as $AA$.  
Issues related to multiple scattering, higher twist effects, and changes
in the gluon distribution function can be explored.  
A direct measurement of open charm will
be important if final state recombination of the produced open charm
makes a significant amount of $J/\Psi$'s.

\subsection{The Lifetime and Size of the Matter Produced}

\begin{figure}[htb]
    \begin{center}
        \includegraphics[width=0.50\textwidth]{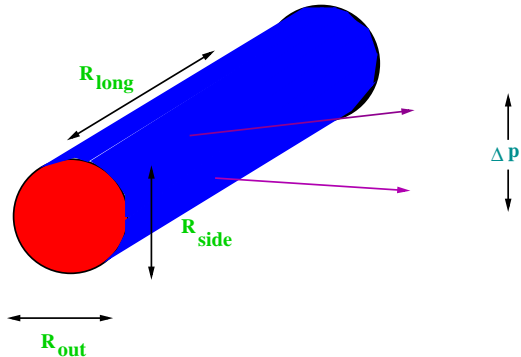}
        \caption{The various radii used for HBT pion interferometry.   
                            }\label{hbt1}   
 \end{center}
\end{figure}

The measurement of correlated pion pairs, the so called HBT pion
interferometry, can measure properties of the space-time volume from
which the hadronic matter emerges in heavy ion collisions.\cite{hbtreview}  
The quantities
$R_{long}, R_{side}$ and $R_{out}$ shown in Fig. \ref{hbt1} measure the
transverse size of the matter at decoupling and the decoupling time.
\begin{figure} [htb]
   \centering
       \mbox{{\epsfig{figure=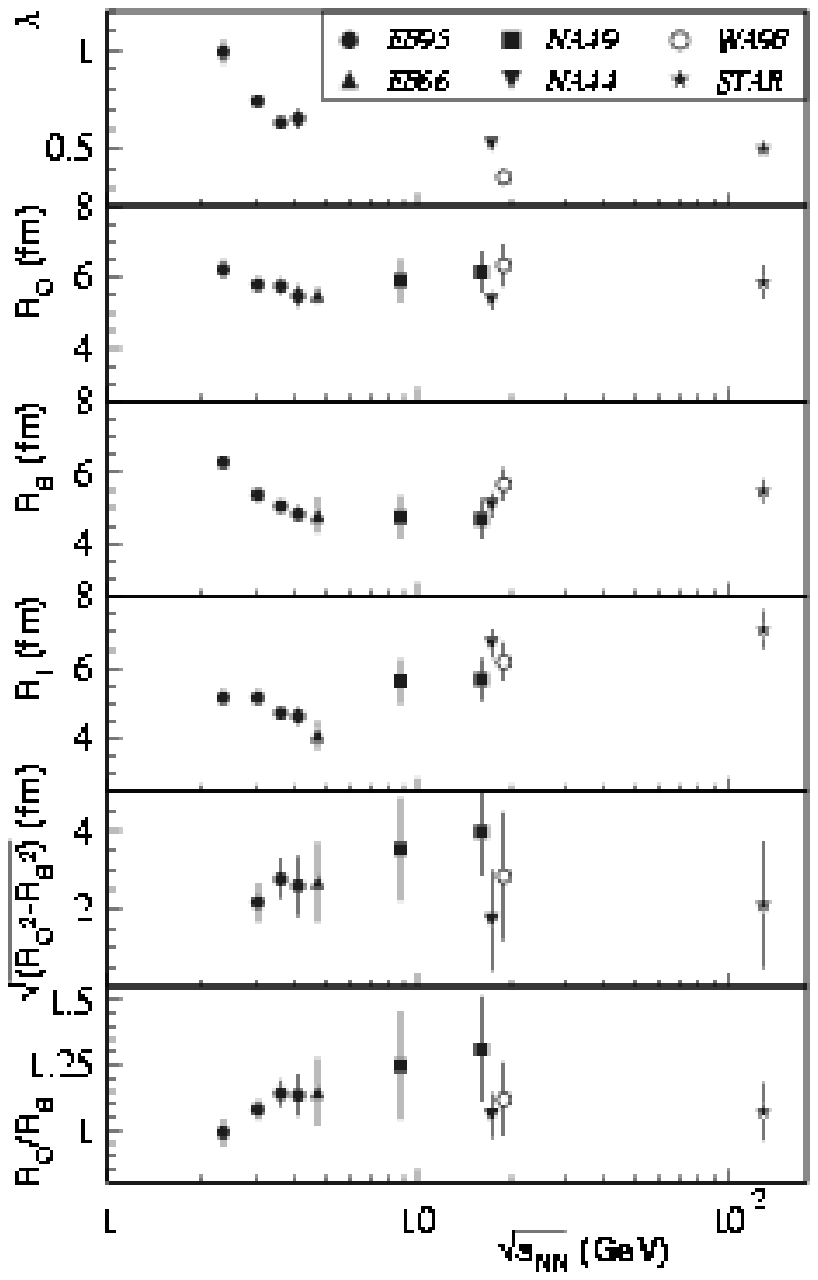,
        width=0.50\textwidth}}\quad
             {\epsfig{figure=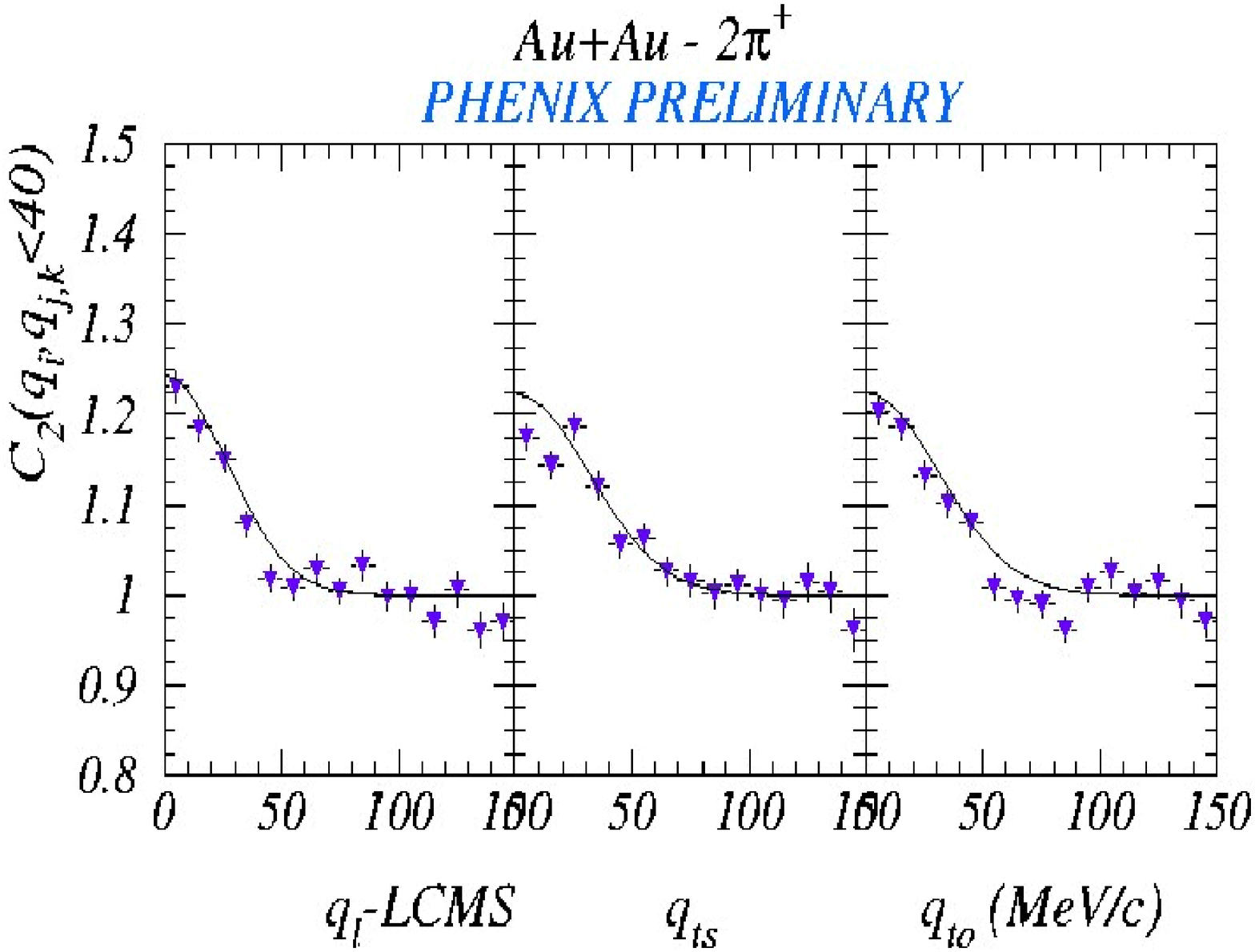,
        width=0.50\textwidth}}}
        \caption{a) The various HBT radii measured in heavy ion 
experiments including the new data from STAR.  b)
The correlation functions which determine the radii as a function of the
pair momenta measured in PHENIX.}
        \label{hbt2}
\end{figure}

In Fig. \ref{hbt2}, the data from STAR and PHENIX 
is shown.\cite{hbtstar}-\cite{hbtphenix}  There is only a 
weak dependence on energy, and the results seem to be more or less what
was observed at CERN energies.  This is a surprise, since a longer time
for decoupling is expected at RHIC.  Perhaps the most surprising result is
that $R_{out}/R_{side}$ is close to 1, where most theoretical expectations give
a value of about $R_{out}/R_{side} \sim 2$.\cite{heinzhbt}-\cite{soffhbt}
Perhaps this is due to greater than expected opacity of the emitting 
matter?   At this time, there is no consistent theoretical description
of the HBT data at RHIC.  
Is there something missing in our space-time picture?

\subsection{The Flavor Composition of the Quark Gluon Plasma}

The first signal proposed for the existence of a Quark Gluon Plasma in heavy
ion collisions was enhanced strangeness production.\cite{muller}
This has lead to a comprehensive program in heavy ion collisions to measure 
\begin{figure} [htb]
   \centering
       \mbox{{\epsfig{figure=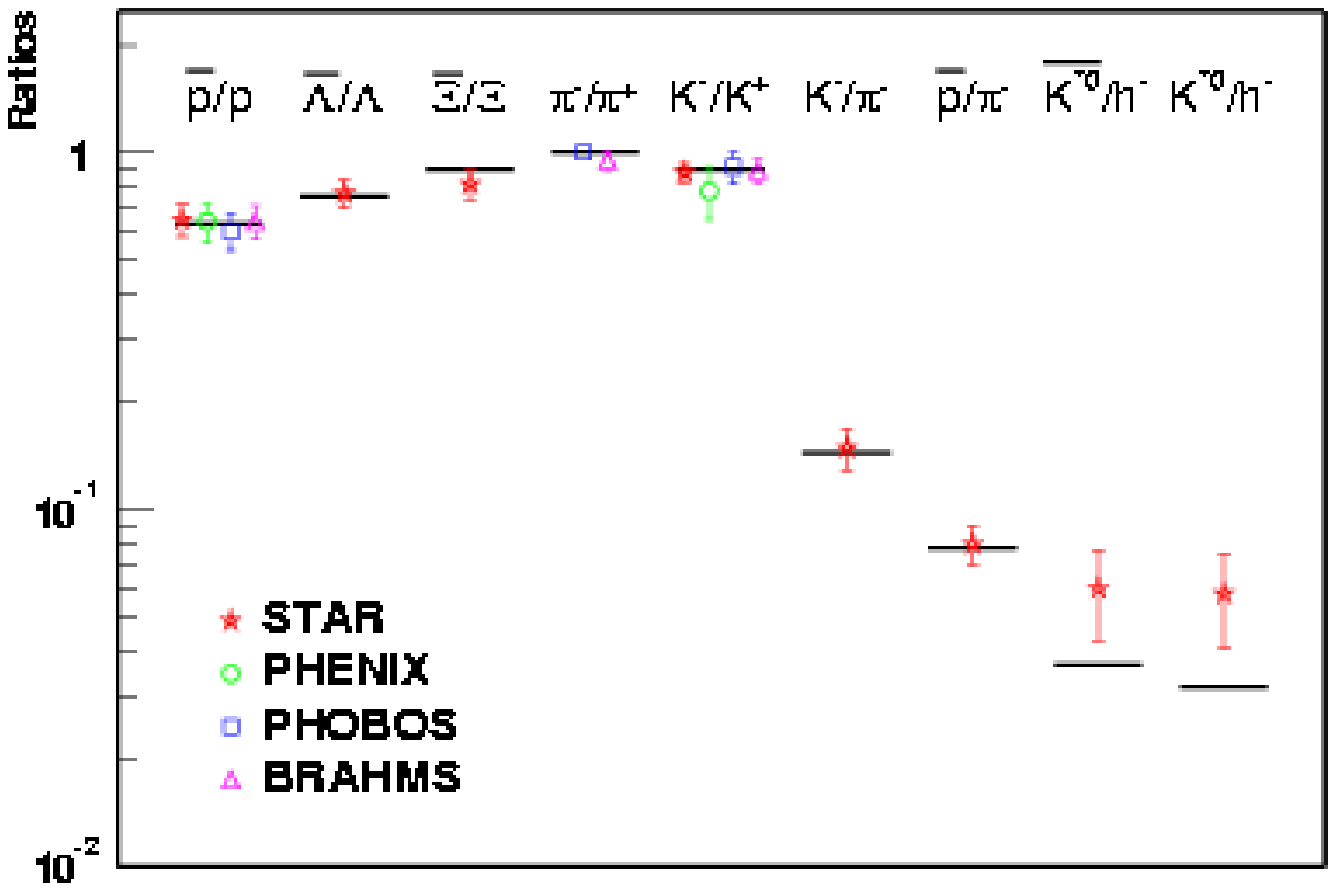,
        width=0.50\textwidth}}\quad
             {\epsfig{figure=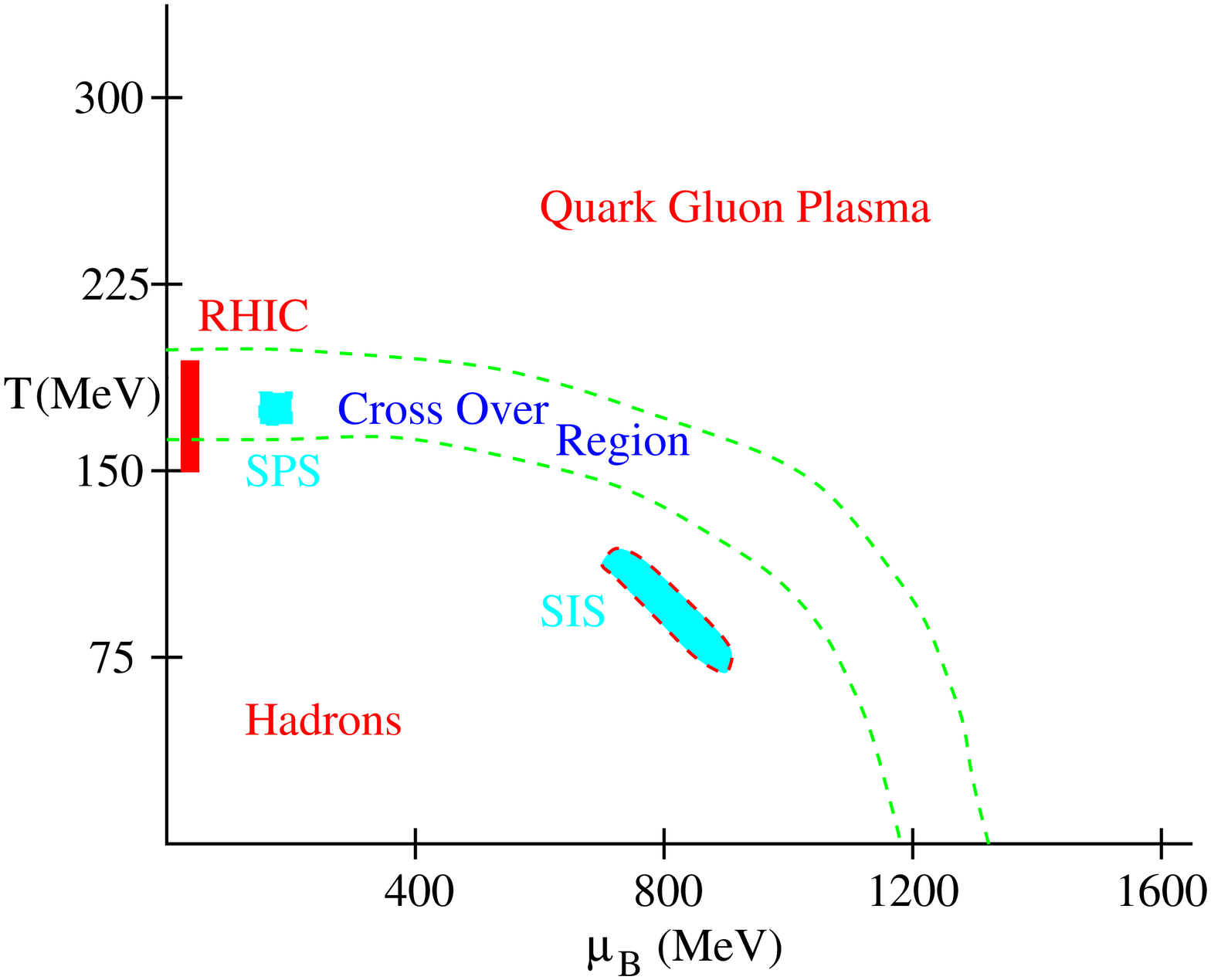,
        width=0.50\textwidth}}}
        \caption{a) Various ratios of particle abundances and
the RHIC data.  The lines are the predictions of a thermal model.  b) 
The temperature vs baryon chemical potential for a thermal model
which is fit to data at various energies.  The dashed line is a 
hypothetical phase boundary between a Quark Gluon Plasma and a hadronic gas.}
        \label{chem}
\end{figure}
the ratios of abundances of various flavors of particles.\cite{nuxu}.
In Fig. \ref{chem}a, the ratios of flavor abundances is compared
to a thermal model for the particle 
abundances.\cite{cleymans} - \cite{gorenstein1}  The fit is quite good.
In Fig. \ref{chem}b, the temperature and baryon chemical potential 
extracted from these fits is shown for experiments at various energies
and with various types of nuclei.  It seems to agree reasonably well
with what might be expected for a phase boundary between hadronic
matter and a Quark Gluon Plasma.

This would appear to be a compelling case for the production of a 
Quark Gluon Plasma.  The problem is that the fits done for heavy ions to
particle abundances work even better in $e^+e^-$ collisions.  One definitely
expects no Quark Gluon Plasma in $e^+e^-$ collisions.  There is a deep
theoretical question to be understood here:  How can thermal models
work so well for non-thermal systems?  Is there some simple saturation of
phase space?  The thermal model description can eventually be made compelling
for heavy ion collisions once the degree of thermalization in
these collisions is understood.

\section{Summary}

\begin{itemize}

\item{\bf What are we trying to understand?}

We want to make and study new forms of high energy density matter.
At early times in a heavy ion collisions, this might be a Color
Glass Condensate, and at later times a Quark Gluon Plasma.

\item{\bf What have we already learned?}

Matter has been produced at energy densities so high it can only be simply
described in terms of quarks and gluons.  This matter is strongly interacting
with itself.  The multiplicity distributions of matter produced in heavy ion
collisions is consistent with it being a Colored Glass.  Flow and $p_T$
distributions are consistent with both a Quark Gluon Plasma and a Colored 
Glass.

\item{\bf What do we expect to learn?}

By measuring the $p_T$ distributions of heavy hadrons, we can resolve the 
differences between various theoretical descriptions of the evolution
of the produced matter.  By measuring high $p_T$ hadrons in AA and pA
collisions, we can determine whether jet quenching occurs, and to what degree
the matter has thermalized.

If thermalization has occurred, we can then perhaps measure the equation
of state.  This is important because the equation of state reflects the
bulk properties of the matter, and determines hydrodynamic and static
properties of Quark Gluon Matter.

\item{\bf What do we hope to learn?}

Through measurements of lepton pairs, we can determine whether or not
resonances have shifted in mass, have broadened or melted.  
Through measurements of pion pairs, we can find the spatial size and 
lifetime of the matter produced in heavy ion collisions, and test space-time
pictures of ultrarelativistic nuclear collisions.  Flavor abundances can
tell us about the macroscopic properties of hadronic matter.  The $J/\Psi$
abundance may tell us whether the matter produced in these collisions is 
confining or not.

\end{itemize}

{\bf 
We must turn to the problem of determining the properties of
matter made in heavy ion collisions.}  We have made matter at such a high
energy density that it can only be simply described in terms of
quarks and gluons.  The future will be to understand the properties of 
this matter.

At present we are developing the theoretical and experimental tools
to analyze ultra-relativistic heavy ion collisions.  It is a learning time
for both theorists and experimentalists.  Ultimately, we will need experiments
with $pp$, $pA$ and $AA$ to resolve the initial state effects due
to a Color Glass Condensate from effects related to the evolution of the matter
as a Quark Gluon Plasma.  A more precise determination of the properties
of Color Glass Condensate would be possible with a high energy 
$eA$ collider or in the LHC heavy ion experiments.

\section{Acknowledgments}

I thank my colleagues in the theory groups at BNL who generously
helped me in the development of this presentation. 

This manuscript has been authorized under Contract No. DE-AC02-98H10886 with
the U. S. Department of Energy.


\begin{thebibliography} {99}

\bibitem{earlywork}  J. C. Collins and M. J. Perry, {\it Phys. Rev. Lett.}
{\bf 34}, 1353 (1975); G. Baym and S. A. Chin, {\it Phys. Lett.} {\bf B62},
241 (1976).

\bibitem{earlymc} L. McLerran and B. Svetitsky,
{\it Phys. Lett.} {\bf B98}, 195 (1981); {\it Phys. Rev.}
{\bf D24}, 450 (1981); J. Kuti, J. Polonyi, and K. Szlachanyi,
{\it Phys. Lett.} {\bf B98}, 199, (1981); J. Engels, F. Karsch and H. Satz,
{\it Phys. Lett.} {\bf B101}, 89 (1981).

\bibitem{pd} B. Svetitsky and L. Yaffe,
{\it Phys. Rev.} {\bf D26}, 963 (1982); {\it Nucl. Phys. } {\bf B210},
423 (1982).
 R. Pisarski and F. Wilczek, {\it Phys. Rev. } {\bf D29},
338 (1984);

\bibitem{colorsuper} R. Rapp, T. Schafer E. Shuryak and M. Velkovsky,
{\it Phys. Rev. Lett.} {\bf 81}, 53 (1998)
M. Alford, K. Rajagopal and 
F. Wilczek,  {\it Phys. Lett.} {\bf B422}, 247 (1998).

\bibitem{recentmc} For a summary of recent results see the excellent 
review talk by F. Karsch, Lectures at the 40'th Internationale
Universitatswochen fuer Theoretische Physik, Dense Matter,
Schladming, Styria, Austria, 3-10 March 2001, hep-lat/0106019.

\bibitem{cgc}  L. V. Gribov, E. M. Levin and M. G. Ryskin, {\it Phys. Rept.}
{\bf 100}, 1 (1983); A. H. Mueller and Jian-wei Qiu, {\it Nucl. Phys.}
{\bf B268}, 427 (1986); L. D. McLerran and R. Venugopalan, {\it
Phys. Rev. } {\bf D49}, 2233(1994); 3352 (1994); E. Iancu, A. Leonidov
and L. D. McLerran, {\it Nucl. Phys.} {\bf A692}(2001); 
E. Ferreiro E. Iancu, A. Leonidov
and L. D. McLerran, hep-ph/0109115; 

\bibitem{aa} A. Kovner, L. D. McLerran and H. Weigert, {\it Phys. Rev. }
{\bf D52}, 6231 (1995); 3809 (1995); A. Krasnitz and R. Venugopalan,
{\it Phys. Rev. Lett} {\bf 84}, 4309 (2000); {\it Nucl. Phys. } {\bf B557},
237 (1999); A. Krasnitz, Y. Nara and R. Venugopalan, {\it Phys. Rev. Lett.}
{\bf 87 }, 192302 (2001).

\bibitem{bjorken} J. D. Bjorken, {\it Phys. Rev. } {\bf D27}, 140 (1983).


\bibitem{star}  C. Adler et. al. {\it Phys. Rev. Lett.} {\bf 87},
112303 (2001)

\bibitem{phenix} K. Adcox et. al. {\it Phys. Rev. Lett.} {\bf 86}, 3500
(2001); {\bf 87}, 052301 (2001).

\bibitem{phobos} M. D. Back et. al. {\it Phys. Rev. Lett.} {\bf 85},
3100 (2000); {\bf 88}, 22302 (2002).

\bibitem{brahms} J. Beardon et. al. {\it Phys. Lett. } {\bf B523}, 
227 (2001)

\bibitem{kn} D. Kharzeev and M. Nardi, {\it Phys. Lett} {\bf B507}, 121 (2001).

\bibitem{npart} B. Back et. al. nucl-ex/0105011; K. Adcox et. al. 
{\it Phys. Rev. Lett.} {\bf 86}, 3500 (2001).

\bibitem{hijing} M. Gyulassy and Xin-Nian Wang, {\it Comput. Phys. Commun}
{\bf 83}, 307 (1994).

\bibitem{ekrt} K. Eskola, K. Kajantie, P. Ruuskanen and K. Tuominen, 
{\it Nucl. Phys. } {\bf B570}, 379 (2000); K. Eskola, K. Kajantie and K.
Tuominen, {\it Phys. Lett.} {\bf B497}, 39 (2001)

\bibitem{kl} D. Kharzeev and E. Levin, nucl-th/0108006.

\bibitem{phbr}   B. Back et. al. {\it Phys. Rev. Lett.} 
{\bf 87}, 102303 (2001); I. Beardon et. al. 
Nucl-ex/0112001.

\bibitem{flowth}  S. Voloshin and Y. Zhang, {\it Z. Phys. } 
{\bf C70}, 665 (1996); A. M. Poskhanzer and S. A. Voloshin, {\it Phys. Rev. } 
{\bf C58}, 1671 (1998); J. Y. Ollitrault, {\it Phys. Rev.} {\bf D46},
229 (1992)

\bibitem{flowstar} K, H. Ackermann et. al. {\it Phys. Rev. Lett.}
{\bf 86}, 402 (2001); C. Adler et. al. {\it Phys. Rev. Lett.} {\bf 87},
182301 (2001).

\bibitem{flowphobos} R. Lacey (for the PHENIX collaboration),
{\it Nucl. Phys. } {\bf A698}, 559 (2002)

\bibitem{heinzkolb} Peter F. Kolb, J. Sollfrank, and U. Heinz,
{\it Phys. Lett. } {\bf B459}, 667 (1999); P. F. Kolb, P. Huovinen,
U. Heinz and H. Heiselberg, {\it Phys. Lett.} {\bf B500}, 232 (2001). 

\bibitem{rajualex} A. Krasnitz, Y. Nara and R. Venugopalan,
in preparation.

\bibitem{starjet} C. Adler, {\it Phys. Rev. Lett.} {\bf 87}
112303 (2001)

\bibitem{phenixjet} K. Adcox et. al. {\it Phys. Rev. Lett} {\bf 88}, 022301
(2002).

\bibitem{schaffner}  J. Schaffner-Bielich, D. Kharzeev, L. D. McLerran and
R. Venugopalan, nucl-th/0108048.

\bibitem{shuryak} D. Teaney and E. V. Shuryak, {\it Phys. Rev. Lett.} 
{\bf 83}, 4951 (1999): D. Teaney, J. Lauret and E. V. Shruyak, 
nucl-th/0110037


\bibitem{srivastava} D. K. Srivastava, {\it Phys. Rev.} {\bf C64},
064901 (2001). 

\bibitem{ceres} G. Agakishiev et. al. {\it Nucl. Phys.} {\bf A638}, 159 (1998).

\bibitem{brown} G. E. Brown and M. Rho, {\it Phys. Rept.}, {\bf 269}, 
333 (1996).

\bibitem{kapusta} J. Kapusta, D. Kharzeev, L. D. McLerran, {\it Phys. Rev.}
{\bf D53}, 5028 (1996).

\bibitem{rapp} R. Rapp, G. Chanfry and J. Wambach,
{\it Phys. Rev. Lett.} {\bf 76}, 368 (1996).

\bibitem{na50}For the latest results, see M. C. Abreau et. al.
{\it Nucl. Phys. } {\bf A661}, 93 (1999).

\bibitem{hadab} J. Geiss, E. Bratskaya, W. Cassing and C. Greiner, 
nucl-th/981005,; C. Spieles, R. Vogt, L. Gerland, S. A. Bass, M. Bleicher, 
H. Stocker and W. Greiner, {\it Phys. Rev. } {\bf C60}, 054901 (1999);
D. E. Kahana and S. H. Kahana, {\it Phys. Rev. } {\bf C60}, 065206 (1999);
N. Armesto, A. Capella, E. Ferreiro, A. Kaidalov and D. Sousa,
{\it Nucl. Phys. } {\bf A698}, 583 (2002).

\bibitem{matsui} T. Matsui and H. Satz, {\it Phys. Lett.}, {bf B178},
416 (1986).

\bibitem{kharzeev} D. Kharzeev and H. Satz, {\it Phys. Lett.} {\bf B334},
155 (1994).

\bibitem{blaizot} Jean-Paul Blaizot and Jean-Yves Ollitrault, 
{\it Phys. Rev. Lett.} {\bf 77}, 1703 (1996).

\bibitem{capella} J. Armesto and A. Capella, {\it Phys. Lett. }
{\bf B430}, 23 (1998); A. Capella, E. G. Ferreiro, A. Kaidalov,
{\it Phys. Rev. Lett.} {\bf 85}, 2080 (2000).

\bibitem{qiu} Jian-wei Qiu, James P. Vary and Xiao-fei Zhang, hep-ph/9809442.

\bibitem{rafelski} R. Thews, M. Schroeder and J. Rafelski,
{\it Phys. Rev. } {\bf C63}, 054905 (2001).

\bibitem{stachel} P. Braun-Munzinger and J. Stachel, {\it Phys. Lett.}
{\bf B490}, 196 (2000).

\bibitem{redlich} P. Braun-Munzinger and K. Redlich, {\it Eur. Phys. J.}
{\bf C16}, 519 (2000).

\bibitem{gorenstein} M. Gorenstein and M. Gazdzicki,
{\it Phys. Rev. Lett.} {\bf 83}, 4009 (1999).
M. Gorenstein, A. P. Kostyk, H. Stoecker and W. Greiner,
{\it Phys. Lett. } {\bf B509}, 277 (2001); M. Gorenstein, A. Kostyk, 
L. McLerran, H. Stoecker and W. Greiner, hep-ph/0012292; 
M. Gorenstein, A. Kostyk,
H. Stocker and W. Greiner, {\it Phys. Lett.} {\bf  B524}, 265 (2002)

\bibitem{hbtreview} M. Gyulassy, S. Kauffmann and L. Wilson,
{\it Phys. Rev. } {\bf C20}, 2267 (1979).

\bibitem{hbtstar} C. Adler et. al. {\it Phys. Rev. Lett. }
{\bf 87 }, 082301 (2001).

\bibitem{hbtphenix} K. Adcox et. al. nucl-es/0201008.

\bibitem{heinzhbt} S. Chapman, P. Scotto and U. Heinz, {\it Phys. Rev. Lett.}
{\bf 74}, 4400 (1995); S. Chapman and U. Heinz, {\it Phys. Lett}
{\bf B340}, 250 (1994).

\bibitem{soffhbt} S. Soff, S. Bass and A. Dumitru,
{\it Phys. Rev. Lett.} {\bf 86}, 3981 (2001).

\bibitem{muller} B. Muller and J. Rafelski, {\it Phys. Rev. Lett.}
{\bf 48} 1066 (1986); P. Koch, B. Muller and J. Rafelski,
{\it Phys. Rept.} {\bf 142} 167 (1986).

\bibitem{nuxu}  For a summary see:
M. Kaneta and N. Xu, {\it J. Phys. } {\bf G27}, 589 (2001).

\bibitem{cleymans} For a state of the
art assessment  review, see Jean. Cleymans, hep-ph/0201142; 
J. Cleymans and K. Redlich, {\it Phys. Rev. Lett.} {\bf 81}, 
5284 (1998); {\it Phys. Rev. } {\bf C60}, 054908 (1999).

\bibitem{stachel} P. Braun-Munzinger, J. Stachel, J. P. Wessels and N. Xu,
{\it Phys. Lett.} {\bf B365}, 1 (1996); P. Braun-Munzinger, I. Heppe and
J. Stachel, {\it Phys. Lett. } {\bf B465}, 15 (1999).

\bibitem{gorenstein1} G. Yen and M. Gorenstein,
{\it Phys. Rev.} {\bf C59}, 2788 (1999).



\end{thebibliography}
\end{document}